\documentclass[reprint,
superscriptaddress,
amsmath,amssymb,
aps,
pra,
]{revtex4-2}

\usepackage{graphicx}
\usepackage{dcolumn}
\usepackage{bm}
\usepackage{physics}
\usepackage{ragged2e}
\usepackage{subcaption}
\usepackage{graphicx}
\usepackage{float}
\usepackage{lmodern}
\usepackage{amsmath}
\usepackage{xcolor}
\usepackage{hyperref}
\makeatletter
\def\Hy@appendixstring{Appendix}
\makeatother

\begin{document}

\preprint{APS/123-QED}

\title{Towards Trapped-Ion Thermometry Using Cavity-Based EIT}

\author{Abhijit Kundu}
\affiliation{Department of Physics, Indian Institute of Technology Tirupati, Yerpedu-517619, Andhra Pradesh, India.}

\author{Vijay Bhatt}
\affiliation{Department of Physics, Indian Institute of Technology Tirupati, Yerpedu-517619, Andhra Pradesh, India.}

\author{Arijit Sharma}\email{arijit@iittp.ac.in}
\affiliation{Department of Physics, Indian Institute of Technology Tirupati, Yerpedu-517619, Andhra Pradesh, India.}
\affiliation{Center for Atomic, Molecular, and Optical Sciences and Technologies,
Indian Institute of Technology Tirupati, Yerpedu-517619, Andhra Pradesh, India.}

\date{\today}

\begin{abstract}
We present a technique for measuring ion temperature using cavity-based electromagnetically induced transparency (EIT) applicable for cavity QED systems. This method enables efficient extraction of the ion's phonon occupation number following sub-Doppler cooling close to the motional ground state. The proposed method requires operation in the resolved-sideband regime, where individual motional states can be selectively addressed for all relevant transitions either by selecting appropriate energy levels for the three-level system or by employing strong confinement with high secular frequencies ($\sim 10 MHz$). It relies on monitoring the cavity probe transmission while scanning the probe laser frequency to establish cavity-induced EIT using a control beam, thereby significantly simplifying the measurement procedure. We establish a theoretical model that demonstrates the influence of the thermal state of the trapped ion vis-à-vis the EIT linewidth measured. We show through numerical simulations how the cavity-induced EIT transmission may be used as a thermometry tool to deduce the ion temperature as well as its motional state in the sub-Doppler cooling regime, even for systems that are in the weak coupling regime.
\end{abstract}
\maketitle

\section{Introduction}

\large Trapped ions offer a well-controlled platform for precision metrology, cavity quantum electrodynamics (QED), and quantum information processing owing to their long coherence time along with an exceptional degree of experimental controllability \cite{leibfried2003quantum,Haffner2008PhysRep,Monroe2013Science,vjl6-crbg}.
\newline


\par With successful demonstrations across a wide range of ion species and trapping architectures, resolved sideband cooling has become the workhorse for preparing trapped ions close to the motional ground state \cite{PhysRevLett.62.403,Roos2000PRL,wineland1998experimental}. Achieving and verifying near-ground-state cooling is of central importance, as even a small residual motional excitation can introduce decoherence and significantly reduce the fidelity of quantum operations \cite{wineland1998experimental,RevModPhys.87.1419}. Consequently, precise thermometry following sideband cooling is essential for accurately characterizing the cooling performance and validating the initialization of the motional state prior to subsequent quantum manipulations.
\newline
\par Accurate knowledge of the ion temperature also plays a crucial role in several aspects of trapped-ion quantum technologies. In quantum information processing, precise thermometry helps mitigate motional decoherence and improves the fidelity of entangling gate operations \cite{cirac1995quantum,Blatt2008Nature}. In ion-trap experiments, it enables the characterization of anomalous heating rates and provides insight into noise sources that limit trap performance \cite{RevModPhys.87.1419}. Furthermore, in cavity-QED based quantum network protocols, controlling and diagnosing the motional state of ions is vital for optimizing ion–photon interfaces and ensuring efficient quantum state transfer between stationary and flying qubits \cite{PhysRevA.61.063418,kimble2008quantum}.
\newline
\par Post compensation of excess micromotion, the motion of an ion confined in a radio-frequency Paul trap can be described as a simple harmonic oscillator with discrete, equally spaced energy levels, which define the motional states \cite{wineland1998experimental}. Ion temperature is directly determined by a discrete phonon occupation number that characterizes the motional state. Determining this phonon occupation provides a crucial benchmark for coherent quantum operations and enables a direct evaluation of the ions' thermal state \cite{leibfried2003quantum,wineland1998experimental,PhysRevLett.76.1796}.\\
\par A variety of techniques have been developed to deduce the motional occupation of trapped ions \cite{PhysRevLett.83.4713,PhysRevLett.76.1796,PhysRevA.104.043108}. The most popular method is resolved sideband thermometry, which assumes a thermal motional distribution and uses the ratio of red to blue sideband transition strengths to estimate the mean phonon occupation \cite{PhysRevLett.62.403, PRXQuantum.4.040346}. More generally, reconstruction techniques that do not strictly rely on thermal assumptions are made possible by extracting phonon distribution information from Rabi oscillations in carrier and sideband transitions \cite{PhysRevLett.76.1796,PhysRevLett.83.4713,PhysRevA.104.043108}. These methods are well known and typically require precise internal-state preparation and the detection of projective states.\\
\par Over the past few decades, following the seminal works of Kimble \cite{kimble1998strong}, Haroche \cite{brune1996quantum} and Rempe \cite{reiserer2015cavity}, there has been growing interest in cavity-QED with trapped ions due to its potential for the realization of scalable quantum networks, enhanced light-matter coupling, and effective ion-photon interfaces for quantum information processing schemes and applications \cite{PhysRevLett.110.213605,wang2000cavity,PhysRevA.61.011801,mucke2010electromagnetically,tanji2011vacuum,Duan2010QuantumNetworksIons,Reiserer2022CavityNodes}. Interference-based phenomena such as cavity-assisted electromagnetic-induced transparency (EIT) are made possible by the presence of an optical cavity, which alters the interaction between the ion and the electromagnetic field. Since its discovery, EIT \cite{PhysRevLett.66.2593} has made it possible for a wide range of coherent optical control applications, such as slow light in atomic medium \cite{novikova2012electromagnetically}, quantum memory \cite{ma2017optical}, laser cooling \cite{morigi2000ground}, precision spectroscopy, and sensing \cite{PhysRevLett.110.213605}.
\newline
\par Cavity-based EIT has been studied in a number of regimes to enable features such as narrowing of the cavity linewidth and enhanced optical nonlinearities \cite{mucke2010electromagnetically,tanji2011vacuum,souza2013coherent}. A key feature of cavity-induced EIT is its sensitivity to decoherence mechanisms in the coupled ion-cavity system. Coupling to vibrational sidebands causes phonon-dependent dephasing of the effective $\Lambda$-type system in the presence of motional excitation, changing the coherence and altering the EIT interference \cite{Cirac1992PRA,morigi2000ground}. The cavity-EIT transparency window is thus systematically expanded by thermal phonons. This sensitivity suggests that the cavity-EIT spectrum can provide a spectroscopic signature of the ion's motional temperature.\\
\par Temperature measurements of trapped ions are often performed using resolved-sideband thermometry, which can obtain sensitivity close to the single-phonon level near the motional ground state, as discussed in \cite{leibfried2003quantum}. Ion temperatures in the $10^{-1}-10^{2}$ mK range have also been measured with sub-mK resolution using dark-resonance thermometry, as shown in \cite{rossnagel2015fast}. In trapped-ion systems imaging-based spatial thermometry typically offers precision at the millikelvin level \cite{norton2011millikelvin}. Recently, cavity-based thermometry has been proposed, where the ion temperature is inferred from the sensitivity of the cavity emission to the ion's spatial motion within the cavity mode \cite{vjl6-crbg}. In this method, the thermometry is based on a position-dependent cavity coupling resulting from the ions thermal motion.\\ 

\par In contrast, the thermometry scheme proposed in this article depends on the effective Rabi frequency in a cavity-EIT configuration being modified in a phonon-number dependent manner, which causes a detectable shift in the EIT linewidth. Aa an alternative thermometry method, particularly suited for cavity-QED experiments, our proposed method relies on measuring the cavity transmission spectrum with high frequency resolution. This enables even minute linewidth variations to correspond to discernible temperature changes.\\

\par In this article, we theoretically investigate a cavity-enhanced thermometry scheme for trapped ions that exploits the phonon-induced modification of cavity-induced EIT. We consider a $\Lambda$-type trapped ion cavity-QED system in which an auxiliary internal transition of the ion is driven by a coupling (control) laser and the cavity mode is interacted with a weak probe field. By including vibrational sidebands and the ion's motion coupled to a thermal reservoir in a $\Lambda$-type ion-cavity setup, we demonstrate that the phonon occupation of the trapped ion modifies the cavity EIT linewidth in a systematic manner. This dependence can be used as a direct probe of the mean phonon number by tuning the coupling field to motional sideband transitions. Thus, the suggested method offers a cavity-compatible thermometry technique that does not depend on projective state detection after sideband cooling. Our theoretical framework establishes a direct link between measurable modification of the cavity-EIT resonance and motional decoherence induced by thermal phonon. The proposed approach provides an experimentally feasible alternative to conventional sideband thermometry in trapped-ion cavity-QED systems.\\
\par We show that the coupling laser causes a phonon number-dependent change in the effective Rabi frequency when it is tuned to a motional sideband transition. This results in a systematic and temperature-dependent broadening of the cavity-EIT feature. We determine a quantitative relationship between the measured cavity-EIT linewidth and the ion temperature by numerically solving the complete open-system dynamics, which includes cavity loss, spontaneous emission, and phonon damping caused by a thermal reservoir. Our results indicate that this technique allows for precise thermometry in the resolved-sideband and sub-Doppler regimes and can be expanded to parameter regimes that are consistent with experimentally realized strong coupling ion-cavity systems\cite{mucke2010electromagnetically,tanji2011vacuum,Duan2010QuantumNetworksIons,Reiserer2022CavityNodes}.\\

\par We subsequently extend the model to the multi-ion system by modifying the atom-cavity interaction Hamiltonian to include a summation over all ions coupled to the same cavity mode. In this formulation, each ion interacts with the cavity field identically, and the collective response of the ion ensemble naturally emerges from the full many-body master equation. This approach allows us to examine thermometry in regimes where the cavity response is strengthened by the collective light-matter interaction inspite of the single-ion coupling strength to the intracavity field placing the system in the weak coupling regime.\\
\par To summarize, we establish a direct link between motional decoherence and cavity transmission, and propose a cavity-compatible minimally invasive thermometry technique for trapped ions through modeling and simulations. Beyond thermometry, the demonstrated sensitivity of cavity-EIT to the motional state of the ion creates new opportunities for studying measurement-induced back-action in cavity-QED systems and for monitoring of thermal dynamics.
\newline
\par The article is structured as follows: In \autoref{sec:theory_model}, we demonstrate the theoretical model and Hamiltonian, including the treatment of ion motion and dissipative processes. In \autoref{sec:Transmission Analysis}, numerical results for linewidth extraction and cavity transmission are shown. \autoref{sec:Effect of themal state} examines how phonons affect the cavity-EIT spectrum.  The analysis of ion temperature and the role of thermalization dynamics are discussed in \autoref{sec:Determination of ion temp}. The discussion then extends to multi-ion cavity-QED systems operating in the weak single-ion coupling regime in \autoref{sec:Multi_ion_small_coupling_factor}, where we demonstrate that thermometry is made possible even in cases where individual coupling strengths are small through collective enhancement. \autoref{sec:Multi_ion_large_excited_state_decay} further investigates the applicability of the proposed method to systems with large excited state decay rates. We show conclusively that thermometry remains feasible under realistic experimental conditions. In \autoref{sec-section8}, we evaluate the projected sensitivity of the thermometry method. We present a realistic experimental realization roadmap in \autoref{sec:roadmap} covering parameter regimes compatible with existing ion-cavity platform trap requirements and viable cavity configurations.
Finally, \autoref{sec:conclusion} summarizes the main findings of this work and presents the concluding remarks.
\begin{figure*}[t]
    \centering
    \begin{subfigure}[t]{0.4\textwidth}
        
        \includegraphics[width=\textwidth]{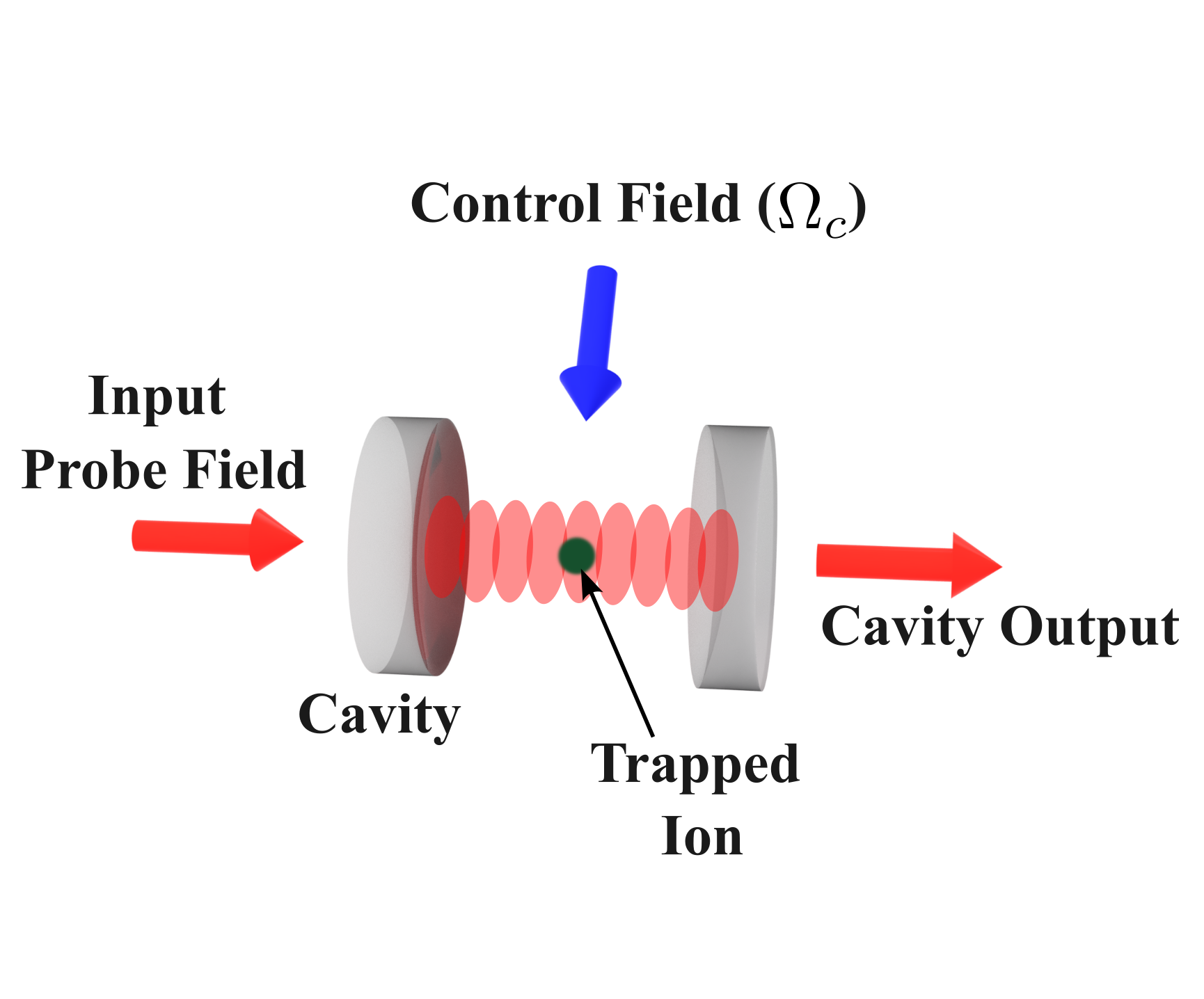}
        \caption{}
        \label{fig1a}
    \end{subfigure}
    \begin{subfigure}[t]{0.4\textwidth}
        
        \includegraphics[width=\textwidth]{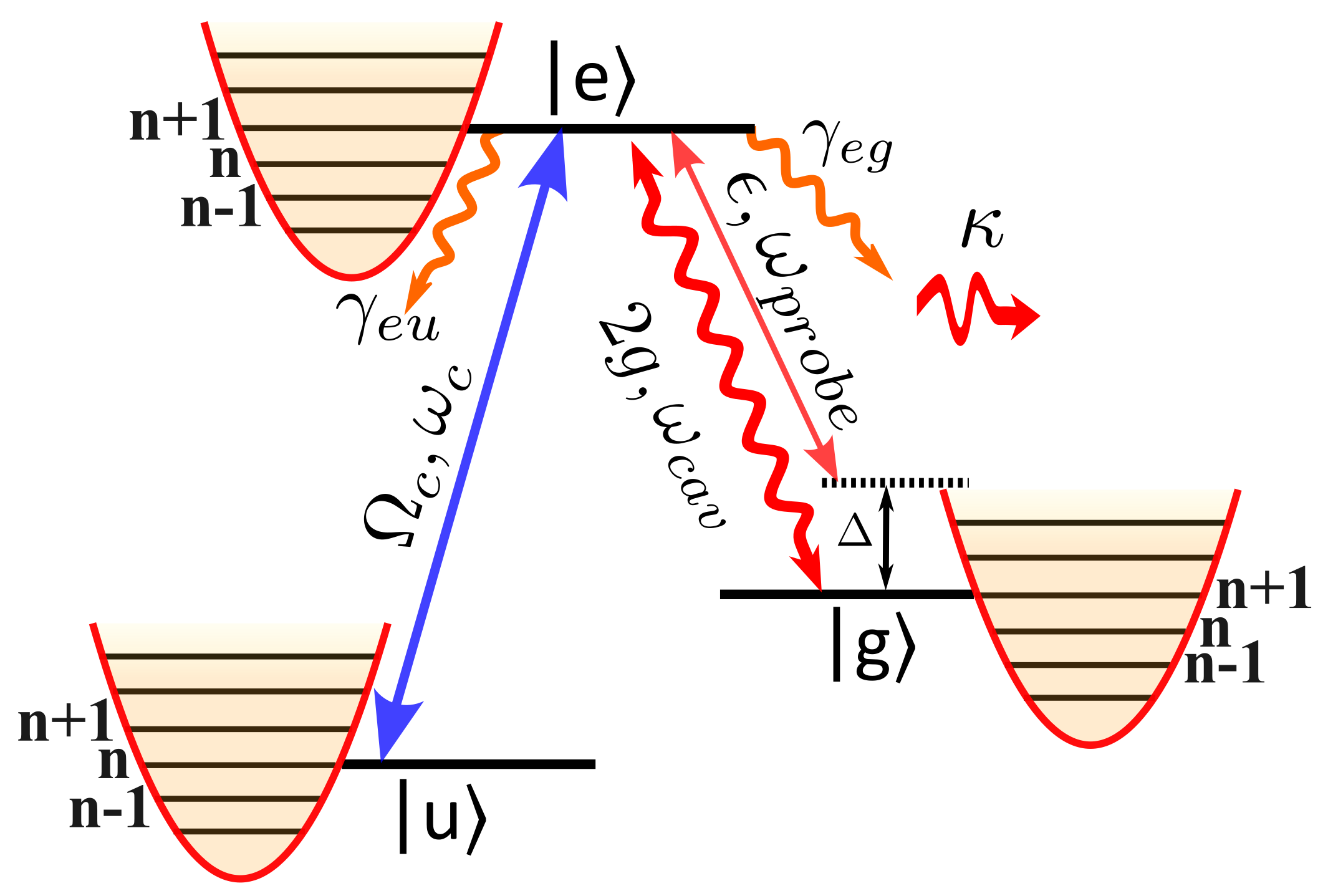}
        \caption{}
        \label{fig:fig1b}
    \end{subfigure}

    \caption{
    \justifying{(a) Schematic of the cavity-based electromagnetically induced transparency (EIT)
    configuration, where a trapped ion interacts with a single cavity mode.
    (b) Energy-level diagram of the cavity-EIT scheme including vibrational states.
    The coupling laser drives the $|u\rangle \rightarrow |e\rangle$ transition with
    the addition of one vibrational quantum, while spontaneous decay to
    $|g\rangle$ preserves the vibrational level. The probe field and cavity mode
    couple the $|e\rangle$ and $|g\rangle$ states.
    }}
    \label{fig:CEIT}
\end{figure*}
\section{Theory and model}\label{sec:theory_model}
\par Cavity-based EIT systems have been extensively studied over the past two decades \cite{mucke2010electromagnetically, tanji2011vacuum, lukin1998intracavity, souza2013coherent, PhysRevLett.100.173602, zhang2007slow, albert2011cavity, liu2017electromagnetically}. These studies demonstrate that the cavity output can be effectively controlled by several key parameters, including the control-field Rabi frequency ($\Omega_c$), the atom-cavity coupling strength ($g_0$), and the cavity decay rate ($\kappa$). 
Among these parameters, $\Omega_c$ represents an experimentally accessible degree of freedom, whereas the others are primarily defined by the cavity geometry, mirror reflectivity, and the choice of atomic energy levels. As a result, parameters such as $g_0$ and $\kappa$ cannot be easily tuned under typical experimental constraints, whereas $\Omega_c$ can be easily modified by tweaking the laser power. As $\Omega_c$ increases, the full width at half maximum (FWHM) of the EIT resonance also increases significantly under a suitable choice of parameters.  This dependence forms the basis of the proposed thermometry scheme. \\

\par For the purpose of ion thermometry, it is essential to include the vibrational levels associated with each relevant electronic energy level. Addressing these vibrational states requires operation in the resolved-sideband regime ($\gamma \ll \omega_{sec}$) \cite{zhang2023sideband, leibfried2003quantum}, where the splitting between adjacent vibrational levels, $\hbar \omega_{sec}$, exceeds the natural linewidth of the corresponding transition, $2\gamma$. Here, $\omega_{sec}$ represents the secular frequency of the trapped ion, and $\gamma$ denotes the spontaneous decay rate of the transition. \\

\par Henceforth, a transition between two energy levels refers to a transition between their corresponding vibrational states. The allowed transitions can be classified into three main types: the red sideband (RSB) transition, the blue sideband (BSB) transition, and the carrier transition, corresponding to $\Delta n = \mp 1$ and $\Delta n = 0$, respectively. 
The vibrational level occupancy, $n$, directly influences the ion-field coupling strength when sideband transitions are addressed.
Consequently, the Rabi frequencies for the RSB and BSB transitions get modified as \cite{leibfried2003quantum},
\begin{equation}
\begin{aligned}
\Omega_{\mathrm{BSB}} &= \eta \Omega_c \sqrt{n+1}, \\
\Omega_{\mathrm{RSB}} &= \eta \Omega_c \sqrt{n}.
\end{aligned}\label{eq:effective_Rabi_freq}
\end{equation}
Where $\eta=kr_0$ is the Lamb-Dicke parameter with $r_0$ being the spread of ground state wavefunction of the ion ($r_0=\sqrt{\hbar/2m\nu}$). $\Omega_c$ is the carrier Rabi frequency and $n$ is the vibrational quantum number.\\ 
\par Unlike the previous case, the Rabi frequency ($\Omega_c$) remains unchanged for the carrier transition. Thus, when the control laser is tuned to the BSB or RSB transition, the effect of the phonon occupancy, $n$, becomes evident in the EIT linewidth through its influence on the Rabi frequency, $\Omega_c$. In other words, variations in the EIT linewidth provide an estimate of changes in the ion's vibrational occupancy. Motivated by this correlation, we define a systematic approach via numerical simulations for determining the ion temperature via phonon occupancy.
\newline
\par We consider a $\Lambda$-type three-level system placed inside a high-finesse optical cavity. The states are considered to be coupled with vibrational states, denoted by $\ket{u,n}$, $\ket{e,n+1}$, and $\ket{g,n}$. The cavity mode interacts with the atomic transition $\ket{g,n+1} \leftrightarrow \ket{e,n+1}$ with coupling strength $g_0$, while a classical control laser couples the transition $\ket{u,n}\leftrightarrow \ket{e,n+1}$ by a change in single quanta of the phonon mode with Rabi frequency $\Omega_{c}$ (see \autoref{fig:fig1b}). A weak input probe field of amplitude $\epsilon$ and frequency $\omega_{probe}$ is applied at probe detuning ($\Delta$). To obtain the cavity output, the $\Delta$ is scanned on both side of the cavity resonance. The spontaneous emission rates from level $|e,n+1\rangle \rightarrow |u,n+1\rangle$ and $|e,n+1\rangle \rightarrow |g,n+1\rangle$ are $\gamma_{eu}$ and, $\gamma_{eg}$ respectively.
\newline

\par We operate in the Lamb-Dicke regime ($\eta \ll 1$), where the control laser and ion's motion can be expanded to first order in the Lamb-Dicke parameter $\eta$. Operators $b$ and $b^{\dagger}$ describe phonon-assisted transitions, achieved by tuning the coupling laser to the first motional sideband of the $\ket{u} \leftrightarrow \ket{e}$ transition \cite{leibfried2003quantum}. Here we consider coherent excitation of the first blue motional sideband. This requires that the motional sidebands are spectrally resolved, i.e., $\omega_{\rm sec}\gg\gamma$, and that the coupling laser is tuned one secular frequency above the carrier transition. Furthermore, the driving strength is chosen such that $\Omega_c\ll\omega_{\rm sec}$, ensuring that the off-resonant carrier excitation probability is approximately $\left(\Omega_c/2\omega_{\rm sec}\right)^2$, and is therefore strongly suppressed. Consequently, only the resonant transition $|u,n\rangle\leftrightarrow|e,n+1\rangle$ is retained in the effective Hamiltonian.\\

Spontaneous processes with $\Delta n\neq0$ are neglected within the first-order Lamb--Dicke approximation. In this regime, transitions accompanied by a change in the motional quantum number are suppressed by factors of order $\eta^2$ relative to carrier decay and therefore constitute higher-order corrections. Since the present work employs the standard first-order sideband Hamiltonian to describe the cavity-EIT dynamics, these higher-order motional processes are consistently neglected. \\

In the rotating frame defined by the probe frequency $\omega_{probe}$, the dynamics of the system can be described by the following time-independent Hamiltonian under the electric dipole and rotating-wave approximation (RWA) (with $\hbar$ =1):

\begin{equation}
\begin{aligned}
H =\;&
\Delta_{p}\sigma_{gg}
- \Delta_{p} a^{\dagger} a
+ g_{0}\left(a^{\dagger}\sigma_{ge} + a\sigma_{eg}\right) \\[4pt]
&+ \eta\Omega_{c}\left(\sigma_{eu} b^{\dagger} + \sigma_{ue} b\right)
+ \epsilon\left(a^{\dagger} + a\right).
\end{aligned}
\label{eq:Ham_with_phonon_1}
\end{equation}

Here, $\sigma_{ij} = \left| i \right\rangle\left\langle j \right|$ ($i,j=u,e,g$) are the ionic transition operators. The detuning parameter $\Delta_{p} = \omega_{probe} - \omega_{eg}$ represents the frequency mismatch between the probe laser and the atomic transition $\ket{g,n+1} \leftrightarrow \ket{e,n+1}$.\\

The first term of the Hamiltonian in \autoref{eq:Ham_with_phonon_1} describes the energy of the state $\ket{g,n+1}$ in the rotating frame, while the second term corresponds to the energy of the cavity photons. The cavity field is represented using the photon annihilation $(a)$ and creation $(a^{\dagger})$ operators.\\

The third term of the Hamiltonian in \autoref{eq:Ham_with_phonon_1} accounts for the coherent interaction between the cavity field and the atomic transition $\ket{g,n+1} \leftrightarrow \ket{e,n+1}$. The fourth term describes the interaction between the atomic transition $\ket{e,n+1} \leftrightarrow \ket{u,n}$ and the vibrational mode of the ion, mediated by the control laser. Here, $b^{\dagger}$ and $b$ are the phonon creation and annihilation operators respectively. The last term represents the coherent drive applied to the cavity field by the pump laser.\\ 

The open dynamics follow the Lindblad master equation \cite{manzano2020short},

\begin{equation}
\begin{aligned}
\frac{d\rho}{dt}
=\;&
-i[H,\rho]
+
\kappa\left(2a\rho a^{\dagger}-a^{\dagger}a\rho-\rho a^{\dagger}a\right) \\[4pt]
&+
\gamma_{eu}\left(2\sigma_{ue}\rho\sigma_{eu}-\sigma_{ee}\rho-\rho\sigma_{ee}\right) \\[4pt]
&+
\gamma_{eg}\left(2\sigma_{ge}\rho\sigma_{eg}-\sigma_{ee}\rho-\rho\sigma_{ee}\right) \\[4pt]
&+\gamma_{b} (n_{\mathrm{th}}+1)
\left(
2b\rho b^{\dagger}
-
b^{\dagger} b\rho
-
\rho b^{\dagger} b
\right) \\[4pt]
&+
\gamma_{b} n_{\mathrm{th}}
\left(
2b^{\dagger}\rho b
-
b b^{\dagger}\rho
-
\rho b b^{\dagger}
\right)
\end{aligned}
\label{master_eq_1}
\end{equation}


where $\rho$ represents the density matrix of the ion-cavity system and $\kappa$ denotes the decay rate of the cavity field. $\gamma_{eg}$ and $\gamma_{eu}$ represent the spontaneous decay rate from state $\ket{e}$ to state $\ket{g}$ and $\ket{u}$. $\gamma_{b}$ is the phonon damping rate and $n_{\mathrm{th}}$ denotes the mean thermal phonon number at equilibrium.\\


The trapped ion is assumed to occupy a thermal motional state after Doppler or sideband cooling. The populations of the harmonic oscillator levels follow the Boltzmann distribution, yielding the mean phonon occupation,
\begin{equation}
    \bar{n} =\frac{1}{\exp(\hbar\omega_{sec}/k_bT)-1} \label{eq:n_avg}
\end{equation}
where $\omega_{sec}$ is the trap secular frequency. Throughout this work, we assume that the ion's motional degree of freedom is in thermal equilibrium with an effective thermal reservoir. Accordingly, the motional state is described by a Gibbs (thermal) distribution over the harmonic oscillator eigenstates, and its mean phonon occupation is given by \autoref{eq:n_avg}. The temperature T therefore refers to the temperature of the effective thermal bath and is inferred from the measured mean phonon occupation.\\

\par The phonon states are inherent to damping due to fluctuating electric field noise present in real ion traps. To account for this scenario, we considered the trapped ion system such that it is coupled to a thermal bath. Fluctuating electric-field noise induces phonon exchange between the ion and its surrounding reservoir, driving the system towards thermal equilibrium. Under this assumption, the mean phonon occupation obeys the standard rate equation for a damped harmonic oscillator coupled to a thermal bath \cite{RevModPhys.87.1419},
\begin{equation}
\dot{\bar{n}}(t) = -\gamma_{b}\,\bar{n}(t) + \gamma_{b}\,n_{th}
\label{eq:nbar_rate}
\end{equation}
where $\gamma_b$ is the motional heating (or damping) rate and $n_{th}$ is the equilibrium mean thermal phonon occupation of the bath. We observe that close to the ground state ($\bar{n}=0$), $\dot{\bar{n}}(t)=\gamma_{b} n_{th}$. This gives an initial estimation of the heating rate of the ion. \autoref{eq:nbar_rate}, also shows that if a long time interval is taken, i.e., at equilibrium, $\bar{n}_{eq}=n_{th}$. So, determining the value of $ n _ {th} $ provides an estimate of the ion temperature owing to \autoref{eq:n_avg}.  


\section{Analysis of Transmission in Cavity-EIT}\label{sec:Transmission Analysis}

To study the system dynamics, the master equation (\autoref{master_eq_1}) has been solved numerically using the Hamiltonian given in \autoref{eq:Ham_with_phonon_1} for a single ion strongly coupled to the intracavity field. The probe laser frequency is scanned across the cavity resonance, keeping the other system parameters fixed. The steady-state density matrix is obtained for each value of the probe detuning. \\

\begin{figure}[htbp]
    
    \includegraphics[width=\columnwidth]{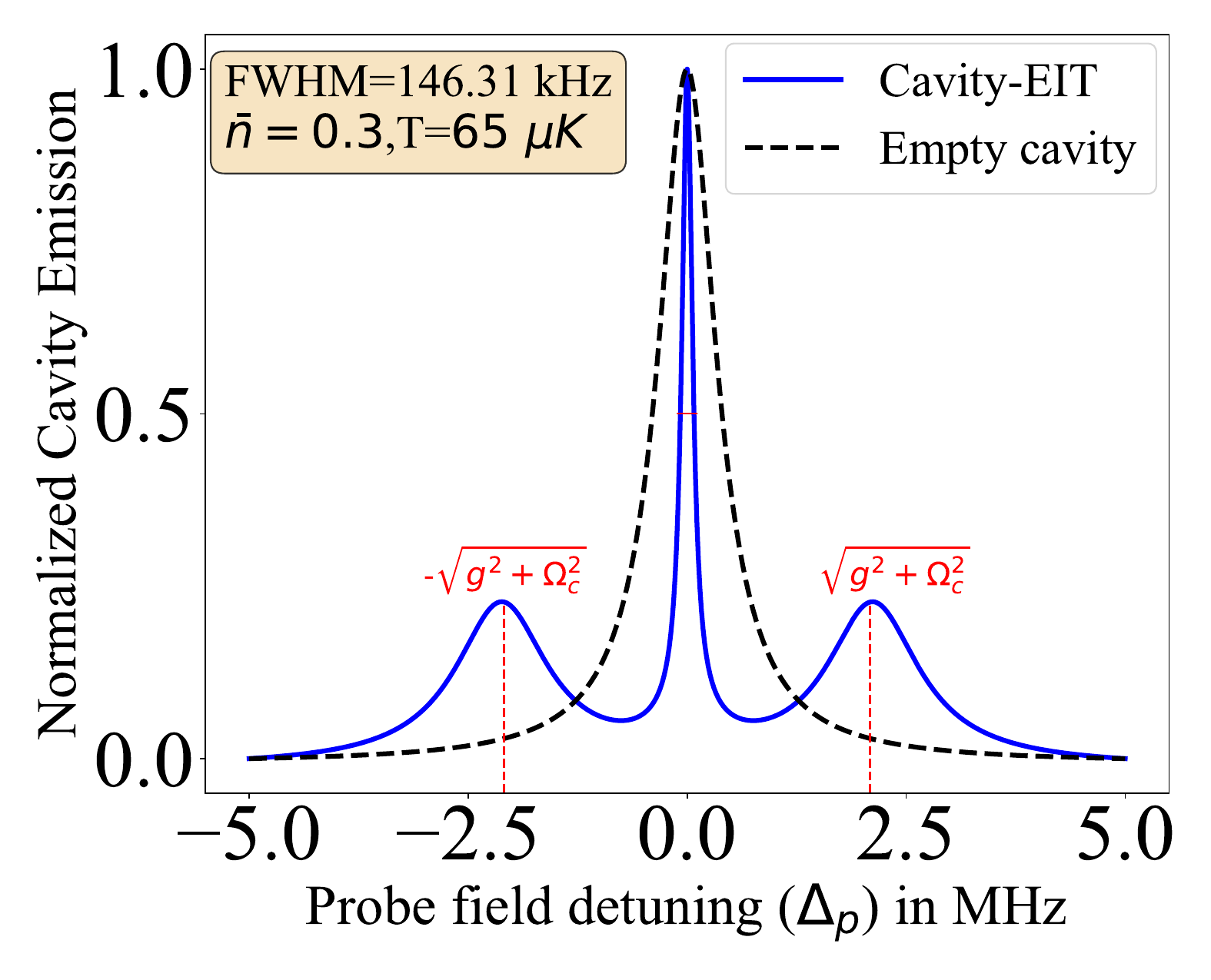}
    \caption{\justifying{
    Simulated cavity-EIT transmission spectrum for 
    $\kappa=0.4\,\mathrm{MHz}$, $g_{0}=5\kappa$, $\Omega_c=3\kappa$ 
    $\gamma_{eg}=\gamma_{eu}=\kappa$, 
    $\bar{n} = 0.3$, and $\gamma_b=0.6\kappa$. 
    The obtained linewidth ($0.15\,\mathrm{MHz}$) is narrower 
    than the empty cavity linewidth $2\kappa=0.8\,\mathrm{MHz}$. 
    The vacuum Rabi splitting peaks occur at 
    $\pm\sqrt{g_0^2 + \eta^2\Omega_c^2} = \pm2.08\,\mathrm{MHz}$. The secular frequency is taken as $2 MHz$ for temperature estimation.
    }}
    \label{fig_single_cav_EIT}
\end{figure}

From the steady-state solution, the inter-cavity photon population is extracted by evaluating the expectation value of the photon number operator ($\langle a^\dagger a \rangle$). The steady-state intra-cavity photon number, which is proportional to the output field leaking through the cavity mirrors, is then used to calculate the transmission spectrum. In \autoref{fig_single_cav_EIT} the resulting transmission profile is illustrated. This spectrum exhibits the characteristic features of previously reported cavity-EIT systems \cite{mucke2010electromagnetically, santos2024fundamental, tanji2011vacuum, souza2013coherent}.\\

\begin{figure*}[t]
    
    \begin{subfigure}[t]{0.4\textwidth}
        
        \includegraphics[width=0.94\textwidth]{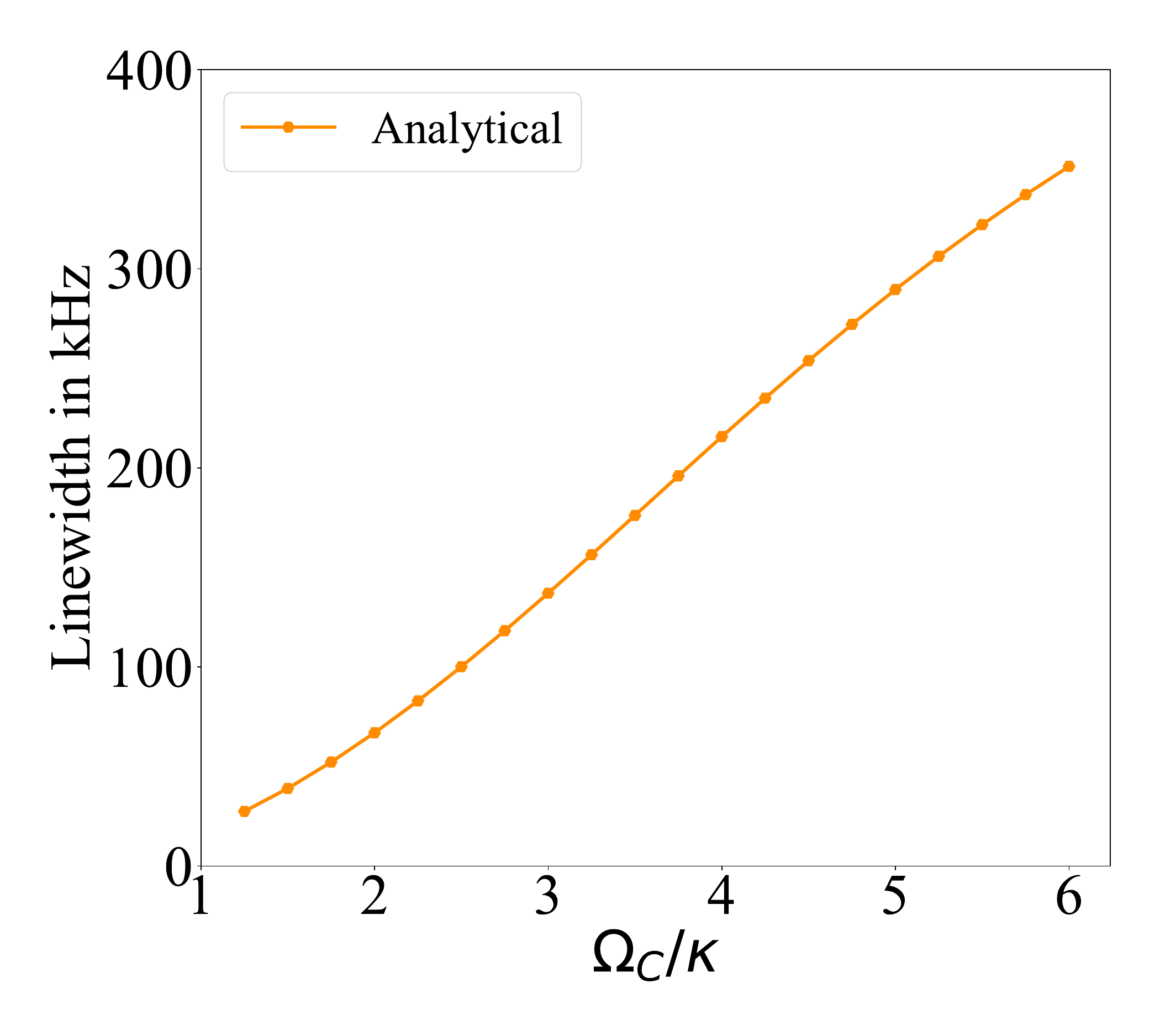}
        \caption{}
        \label{ome_ana}
    \end{subfigure}
    \begin{subfigure}[t]{0.4\textwidth}
        
        \includegraphics[width=\textwidth]{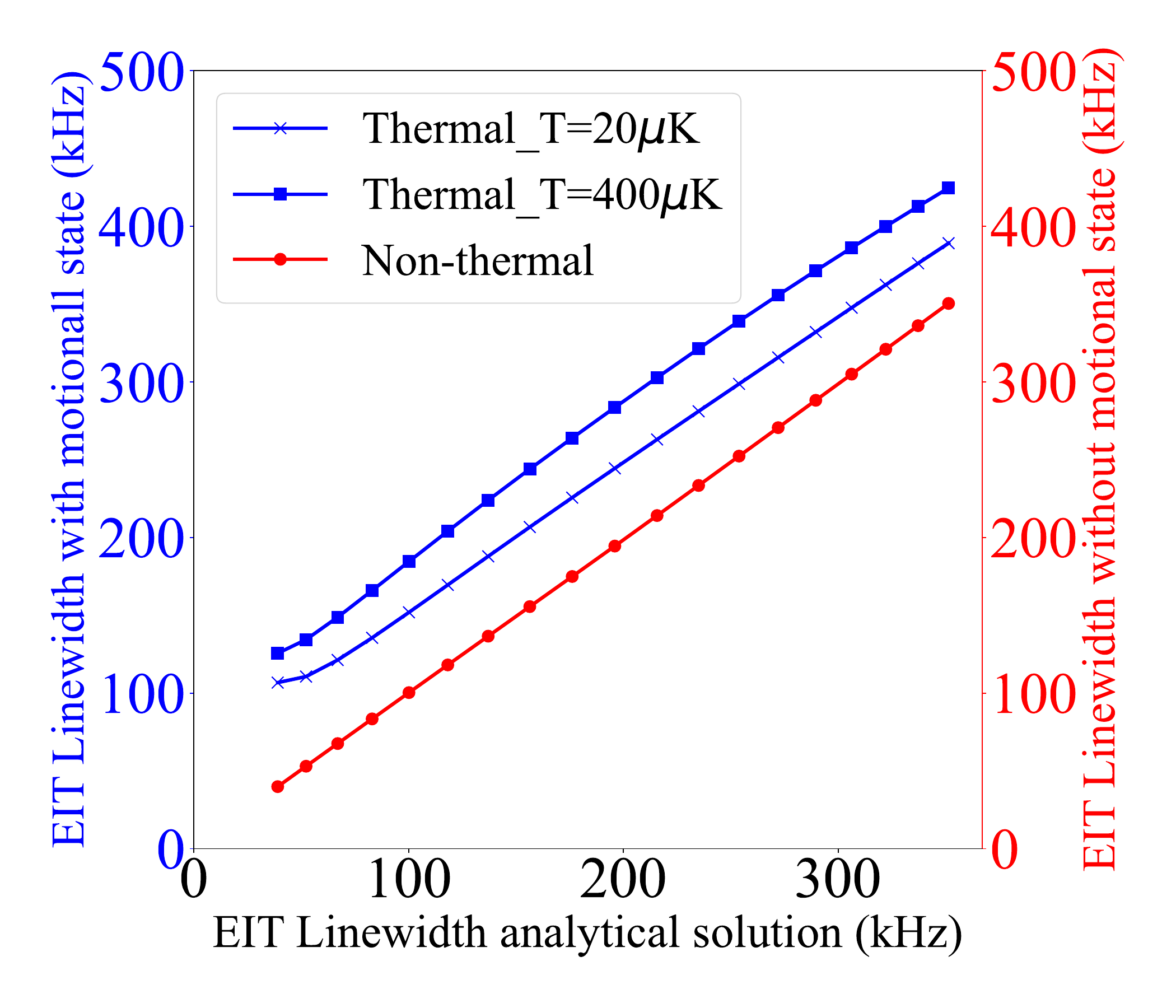}
        \caption{}
        \label{fig:CEIT_b}
    \end{subfigure}

    \caption{\justifying{
    (a) Cavity-EIT linewidth from analytical expression as a function of control field Rabi frequency. (b) Comparison of linewidth without considering motional state vs with motional state for different temperatures of ion. The parameters for the plots chosen as follows, $\kappa=0.4 MHz$, $g_{0}=3\kappa$, $\gamma_{eg}=\gamma_{eu}=\kappa$, $\gamma_b=0.04\kappa$.
    }}
    \label{fig:Effect_of_motional}
\end{figure*}

\par The cavity linewidth is observed to be narrower than that of an empty cavity, and the vacuum Rabi splitting peaks appear at $\Delta = \pm \sqrt{g_{0}^2 + \Omega_c^2}$. For all simulations presented in this work, unless otherwise stated, the system parameters are fixed as follows. All quantities are normalized with respect to the cavity decay rate, $\kappa$, which is taken to be $0.4~\mathrm{MHz}$, a typical value for a high-finesse cavity. The coupling-field Rabi frequency, $\Omega_c$, is set to $2.5\kappa$, while the atom-cavity coupling strength, $g_0$, is chosen as $3\kappa$. The decay rates from the excited states, $\gamma_{ge}$ and $\gamma_{ue}$, are taken to be equal to $\kappa$. In addition, motional-state decoherence is included to account for unwanted heating or cooling of the ion arising from phase noise and radiative fluctuations present in realistic ion traps. To model this effect, a weak coupling between the motional state and the environment is introduced, characterized by a coupling constant $\gamma_b \approx 0.25\kappa$.
\section{Effect of thermal state on cavity EIT linewidth} \label{sec:Effect of themal state}
The usual cavity-EIT formalism discussed and reviewed through multiple theoretical approaches \cite{fleischhauer2005electromagnetically,lukin2003colloquium,dantan2004quantum} as well as experimental investigations \cite{Mucke2010Nature, albert2011cavity} till date has not explicitly considered the ion's motional state in the realization of the cavity-induced EIT phenomenon. This inherently makes an assumption that the ion is fixed in space and the effect of ion thermalization does not affect the cavity-EIT output. When the ion's motion is not considered, the conventional cavity-EIT model Hamiltonian can be written as
\begin{equation}
\begin{aligned}
H_{\mathrm{EIT}} =\;&
\Delta_p\, \sigma_{gg}
- \Delta_p\, a^{\dagger}a
+ g_{0}\left(a^{\dagger}\sigma_{ge} + \sigma_{eg}a\right) \\[4pt]
&+ \Omega_{c}\left(\sigma_{ue} + \sigma_{eu}\right)
+ \varepsilon\left(a^{\dagger} + a\right),
\end{aligned}
\label{eq:Ham_non_thermal}
\end{equation}

where the different terms have their usual meaning as described in \autoref{sec:theory_model}. We refer to this formalism as non-thermal and the Hamiltonian given in \autoref{eq:Ham_non_thermal}, the non-thermal cavity-EIT Hamiltonian. The output of the cavity is traced from the average of the photon number operator with the variation of the probe field detuning. The same cavity transmission can also be analytically calculated from the expression\cite{dissertation_mucke}, 
\begin{equation}
    \mathcal{T}=\frac{\kappa^2}{|(\Delta_p+i\kappa)-\chi|^2}
\end{equation}
where,
\begin{equation}
    \chi=g_{0}^2N\frac{\Delta_p}{(\Delta_p+i(\gamma_{eu}+\gamma_{eg}))\Delta_p-\frac{1}{4}\Omega_c^2}
\end{equation}
$\chi$ represents the atomic susceptibility, and $N$ is the number of atoms coupled to the cavity mode. 
The linewidth depends on the effective Rabi coupling, $\Omega_c$, exhibiting power-broadening behavior. In \autoref{ome_ana}, the relationship between the effective Rabi coupling and the cavity-EIT linewidth is shown using the analytical expression.\\
\par When the motional states are incorporated into the Hamiltonian (\autoref{eq:Ham_non_thermal}), it encompasses a more general description of a trapped ion, the cavity-EIT linewidth deviates from the non-thermal case. As the temperature increases, the thermal nature of the ion leads to a higher motional occupancy. This increase in motional occupation modifies the cavity-EIT linewidth by changing $\Omega_c$, as discussed in detail in \autoref{sec:theory_model}.\\

\par To compare the cavity-induced EIT linewidths in the thermal (when motional states are included) and non-thermal (when motional states are not considered) cases, we perform simulations for both scenarios using the same values of $\Omega_c$ as those employed in the analytical plot in \autoref{ome_ana}. The thermal simulations are carried out over a range of temperatures to capture the dependence of the cavity-induced EIT linewidth on the ion temperature. In \autoref{fig:CEIT_b}, the thermal and non-thermal linewidths are presented, with the analytical linewidths from \autoref{ome_ana} used as the reference on the x-axis.\\  
\par It is shown that, over the entire range, the non-thermal linewidth remains lower than the thermal linewidth and that the linewidth curves shift toward higher values as the temperature increases. This behavior provides further motivation for quantifying the ion temperature, as it modifies the cavity linewidth, which is highly sensitive to temperature-induced changes in the ion’s motional state.

\section{Determination of ion temperature and average phonon number}\label{sec:Determination of ion temp}
We study the cavity-based EIT phenomenon with the motivation of determining the ion temperature. It is evident from \autoref{sec:Effect of themal state} that the linewidth of the cavity-EIT spectrum appears to serve as a probe parameter for the trapped ion temperature. The ion temperature is incorporated in the systems by means of the thermal bath phonon occupancy ($n_{th}$) following \autoref{eq:n_avg} and \autoref{eq:nbar_rate} in steady-state. \\

\par
\begin{figure*}[t]
    
    \begin{subfigure}[t]{0.4\textwidth}
        
        \includegraphics[width=\textwidth]{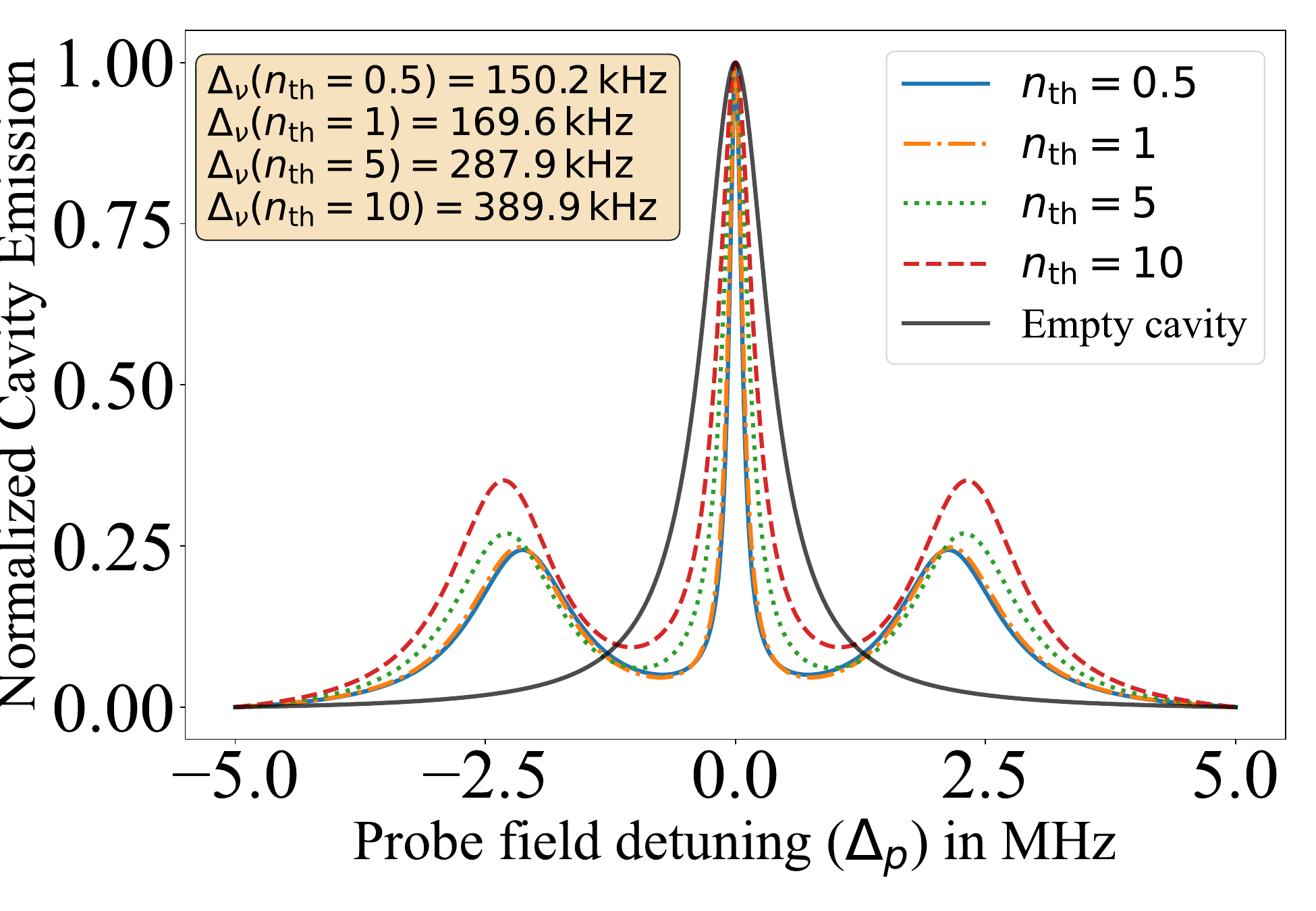}
        \caption{}
        \label{fig:n_bar_temp_linewidth_a}
    \end{subfigure}
    \begin{subfigure}[t]{0.42\textwidth}
        
        \includegraphics[width=\textwidth]{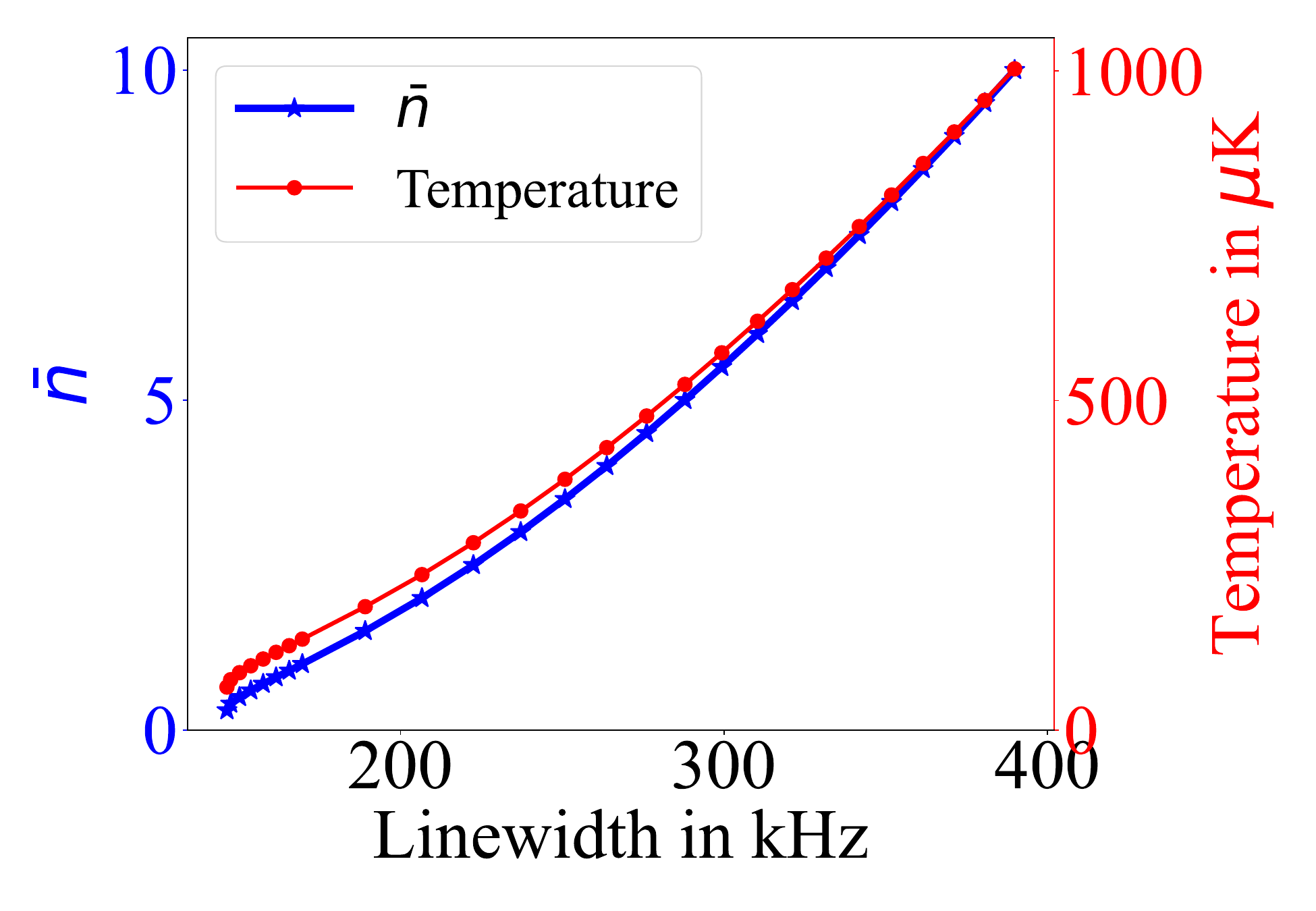}
        \caption{}
        \label{fig:n_bar_temp_linewidth_b}
    \end{subfigure}
    \caption{\justifying{(a) Cavity-EIT spectrum for different $n_{th}$ value (b) Ion temperature and average phonon occupancy ($\bar{n}$) as a function of cavity-EIT linewidth. The cavity decay rate, $\kappa=0.4 MHz$, cavity coupling factor, $g_{0}=5\kappa$.  The decay rates from the excited states are taken as $\gamma_{eg}=\gamma_{eu}=\kappa$. The phonon coupling rate is taken as $\gamma_b=0.6\kappa$.}}
    \label{fig:n_bar_temp_linewidth}
\end{figure*}

 The simulation is performed for four different values of the thermal bath phonon occupancy, $n_{th} = 0.5, 1, 5,$ and $10$, using the same set of parameters described in \autoref{sec:Transmission Analysis}. A clear trend of increasing linewidth with increasing $n_{th}$ is observed (see \autoref{fig:n_bar_temp_linewidth_a}). This behavior indicates that the ion temperature can be mapped to the observed cavity-EIT linewidth. In other words, by analyzing the EIT width of the cavity-EIT spectrum, the ion temperature can be inferred. In the subsequent sections, we discuss the applicability of this technique over a wide range of experimental parameters that govern current and futuristic cavity-QED systems in the strong and weak coupling regimes. \\

\par In addition, the average phonon occupancy, $\bar{n}$, is an important quantity for various applications, such as estimating heating rates \cite{PhysRevA.61.063418}, determining the fidelity of ion trap-based quantum gates \cite{sorensen2000entanglement}, indicating the Lamb-Dicke regime etc \cite{leibfried2003quantum}. Since the phonon occupancy is directly related to temperature (see \autoref{eq:n_avg}), $\bar{n}$ can also be estimated from the cavity-EIT linewidth \cite{Wineland1979RadiativeEffects}. \\

\par We further perform simulations over a wide range of ion temperatures, from which the linewidth is calculated by fitting a Lorentzian function to the spectrum. The average phonon occupancy $\bar{n}$ is extracted by taking the expectation over phonon number operator ($\langle b^\dagger b\rangle$). In \autoref{fig:n_bar_temp_linewidth_b}, the ion temperature and $\bar{n}$ are illustrated as functions of the cavity-EIT linewidth. The predicted temperature exhibits an asymptotic behavior at low linewidth values, indicating that at high temperatures, variations in linewidth are minimal. \\

\par This behavior can be understood from the $\sqrt{n}$ dependence of the control field Rabi frequency, $\Omega_c$, which directly governs the cavity-EIT linewidth. At low temperatures, small changes in temperature lead to rapid variations in $\Omega_c$, as described by \autoref{eq:effective_Rabi_freq}, resulting in noticeable changes in the linewidth. In contrast, at higher temperatures, the linewidth tends to saturate because variations in $\Omega_c$ become negligible. \\

\par The average phonon occupation, ($\bar{n}$), follows the same qualitative trend as the temperature because they are related through the Bose-Einstein relation,
\begin{equation}
    T=\frac{\hbar\omega_{\rm sec}}{k_B\ln\left(1+\frac{1}{\bar n}\right)}
\end{equation}

This exact relation is used throughout the present work to convert between the equilibrium mean phonon occupation and the corresponding bath temperature. In the region of interest of $\bar{n}$, the function exhibits a nearly linear visual trend, although it is not the high-temperature linear approximation.
\\
\par Further, we demonstrate how the ion-cavity coupling strength affects the cavity-EIT linewidth, $g_0$, and Rabi frequency of the coupling laser. We provide linewidth maps in two dimensions for a range of mean thermal phonon number values. Appendix \ref{appen:c} contains the relevant discussion and the requisite plots.
\section{Thermometry of a multi-ion cavity-QED system in the weak coupling regime}\label{sec:Multi_ion_small_coupling_factor}
The discussion so far has focused on a single ion confined in a high-finesse cavity and operating in the strong coupling regime. However, operating in the strong-coupling regime is experimentally challenging with only a single ion, requiring either a near concentric-cavity regime \cite{PRXQuantum.2.020331} or fiber cavities where the polished surfaces are placed very close to each other \cite{PhysRevLett.124.013602}. Interestingly, both of these configurations possess very small mode volumes leading to an increase in the single-atom cooperativity factor in a significant manner. Both regimes of operation are experimentally challenging, with substantial effort necessary in making and aligning cavity mirrors with very high reflectivity for the above two cavity configurations.\\

\par Our analysis reveals that it is possible to perform ion thermometry measurements for multi-ion cavity-QED systems having a relatively low single-ion cavity coupling factor ($g_{0}\ll \kappa, \gamma$) that places such systems in the weak coupling regime ($g_{0}\ll \kappa, \gamma$) for a single ion. We show through numerical simulations that is possible to apply the proposed cavity EIT-based ion-thermometry method discussed earlier even for such systems in the weak coupling regime, albeit using more than one ion coupled to the cavity mode instead of a single one. Using a multi-ion configuration coupled to the cavity mode places the system effectively in the strong coupling regime due to the collective ($g_{eff}\propto\sqrt{N} $) scaling up of the single-ion cavity coupling factor.

The Hamiltonian for a multi-ion cavity-QED system may be stated as

\begin{equation}
\begin{aligned}
H =\;&
\Delta_{p}S_{gg}
- \Delta_{p} a^{\dagger} a
+ g_{0}\left(a^{\dagger}S_{ge} + aS_{eg}\right) \\[4pt]
&+ \eta\Omega_{c}\left(S_{eu} b^{\dagger} + S_{ue} b\right)
+ \epsilon\left(a^{\dagger} + a\right).
\end{aligned}
\label{eq:Ham_with_phonon}
\end{equation}

where, \[
S_{ij} = \sum_{k=1}^{N_{\text{ion}}} \left| i \right\rangle^{(k)} \left\langle j \right|^{(k)}
= \sum_{k=1}^{N_{\text{ion}}} \sigma_{ij}^{(k)} .
\] are the collective raising and lowering ionic operators for $i \neq j$, and ionic energy-level population operators for $i=j$.\\

The system dynamics, including the decoherence and dissipative effects, are determined by the master equation as,

\begin{equation}
\begin{aligned}
\frac{d\rho}{dt}
=\;&
-i[H,\rho]
+
\kappa\left(2a\rho a^{\dagger}-a^{\dagger}a\rho-\rho a^{\dagger}a\right) \\[4pt]
&+
\gamma_{eu} \sum_{k=1}^{N_{\text{ion}}}\left(2\sigma_{ue}^{(k)}\rho\sigma_{eu}^{(k)}-\sigma_{ee}^{(k)}\rho-\rho\sigma_{ee}^{(k)}\right) \\[4pt]
&+
\gamma_{eg}\sum_{k=1}^{N_{\text{ion}}}\left(2\sigma_{ge}^{(k)}\rho\sigma_{eg}^{(k)}-\sigma_{ee}^{(k)}\rho-\rho\sigma_{ee}^{(k)}\right) \\[4pt]
&+\gamma_{b} (n_{\mathrm{th}}+1)
\left(
2b\rho b^{\dagger}
-
b^{\dagger} b\rho
-
\rho b^{\dagger} b
\right) \\[4pt]
&+
\gamma_{b} n_{\mathrm{th}}
\left(
2b^{\dagger}\rho b
-
b b^{\dagger}\rho
-
\rho b b^{\dagger}
\right).
\end{aligned}
\label{eq:Ham_with_phonon2}
\end{equation}

\begin{figure}[htbp]
    \includegraphics[width=\columnwidth]{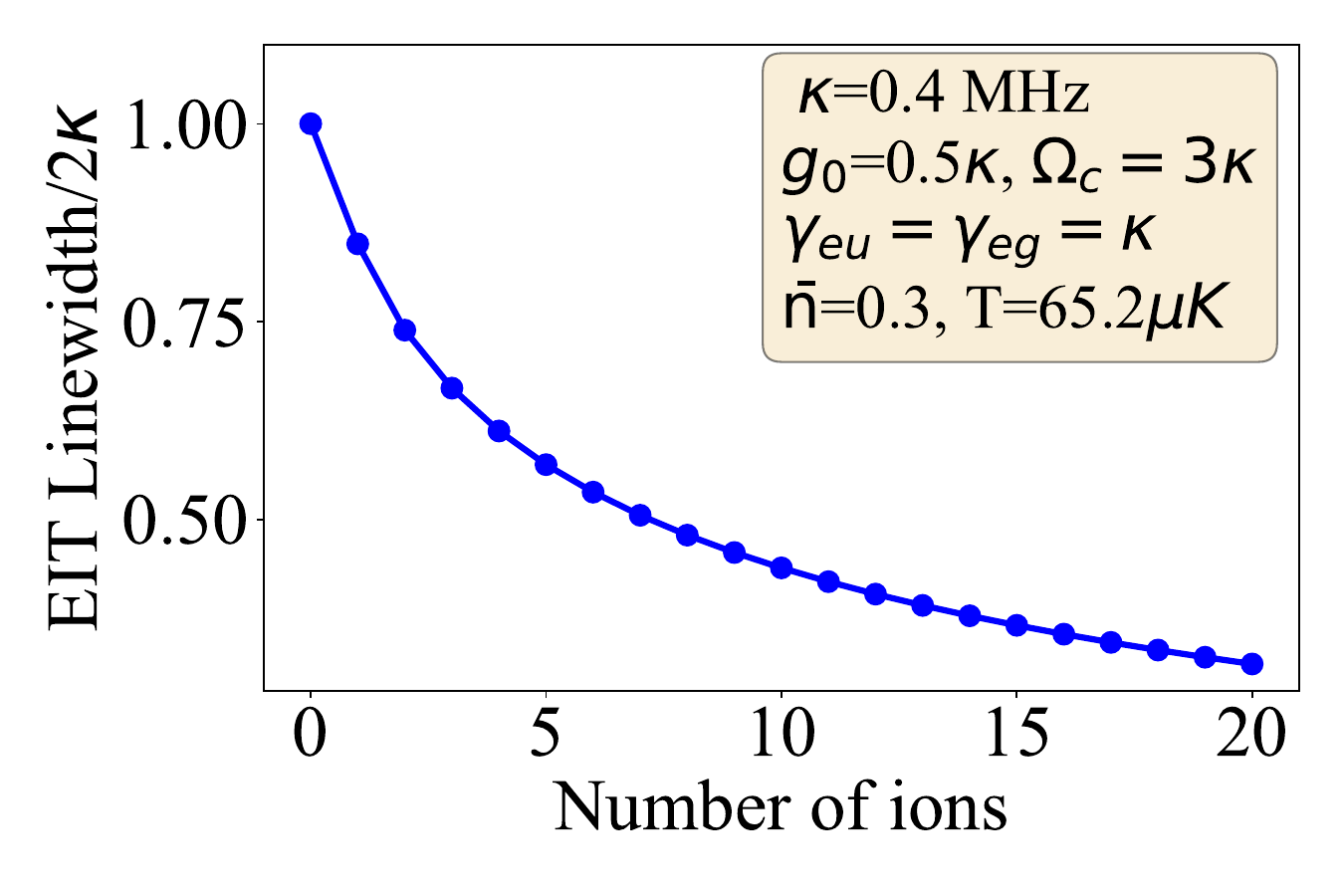}
    \caption{\justifying{
    Cavity-EIT linewidth variation with the number of ions. The simulation parameters are shown in the inset.
    }}
    \label{EIT_width_vs_Nat}
\end{figure}
The method works as one may achieve the collective strong coupling regime with a larger number of ions coupled to the cavity mode, thanks to the scaling of the single-atom cavity coupling factor ($g_{eff}\propto\sqrt{N} $) $g_0$ with ion number, $N$ \cite{PhysRevA.85.023818}. Here $g_0$ is the single-ion cavity coupling factor. The dependence of ion number on cavity-EIT linewidth is depicted in \autoref{EIT_width_vs_Nat}. As the ion number increases, the cavity-EIT linewidth decreases exponentially, making the EIT linewidth even narrower than the single-atom case in the strong coupling regime. It suggests that using a sufficient number of atoms/ions collectively coupled to the cavity, one may be able to realize a very narrow cavity-EIT linewidth, which might be useful for a variety of other applications as well, such as using them as reference transitions in atomic clocks, laser frequency stabilization, etc.\\

\begin{figure}[!h]
    \includegraphics[width=\columnwidth]{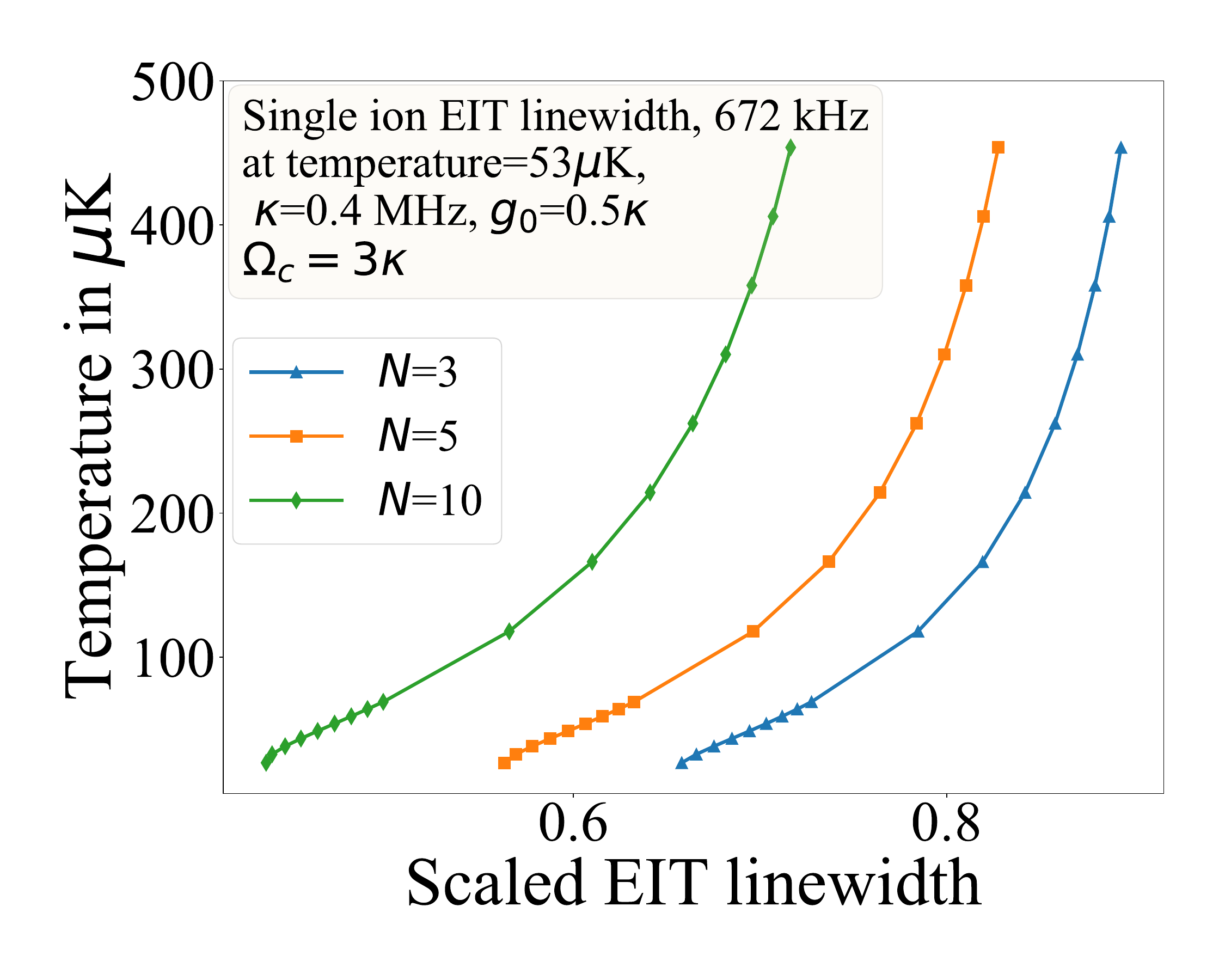}
    \caption{\justifying{
    Temperature vs cavity-EIT linewidth for a multi-ion system. The EIT linewidth is scaled with respect to the empty cavity width. The typical single-ion cavity-EIT linewidth for a chosen temperature is shown in the inset with other system parameters. Except for temperature, the same parameters have been chosen for the simulations performed in this plot. 
    }}
    \label{Linewidth_vs_temp_for different_Nat}
\end{figure}

In the multi-atom/ion cavity system, the ion-cloud temperature may be estimated in a similar manner as outlined in the single-ion cavity EIT-based thermometry scheme as discussed in \autoref{sec:Determination of ion temp}. in \autoref{Linewidth_vs_temp_for different_Nat}, we have shown the variation in temperature as a function of cavity-EIT linewidth in a fraction of empty cavity linewidth for different numbers of ions. The variation of the cavity EIT linewidth with temperature indicates that the linewidth is more sensitive to the temperature as one considers a higher number of atoms/ions coupled to the cavity mode. Thus, the proposed thermometry method is particularly well-suited for cavity-QED systems with low ion-cavity coupling strength, especially when  multiple ions are coupled to the cavity mode.\\

In the multi-ion model, we assume that all ions share an identical thermal occupation. This approximation is justified under a uniformly cooled condition, where all the ions are cooled to a sufficiently low temperature ($\approx 10-50 \mu K$) and undergo comparable coupling to the cavity field. Consequently, the collective cavity response can be described using a thermal distribution reminiscent of a single ion.

\section{Thermometry of a multi-ion cavity-QED system with a large excited-state decay rate}\label{sec:Multi_ion_large_excited_state_decay}
The three-level system we have considered for our ion thermometry is an ideal system, with a very low decay rate from the excited state. However, most ion species used in trapped-ion experiments have a broad excited-state linewidth (i.e., a large decay rate), which helps cool the ions efficiently with Doppler cooling.   

\begin{figure}[htbp]
    \includegraphics[width=\columnwidth]{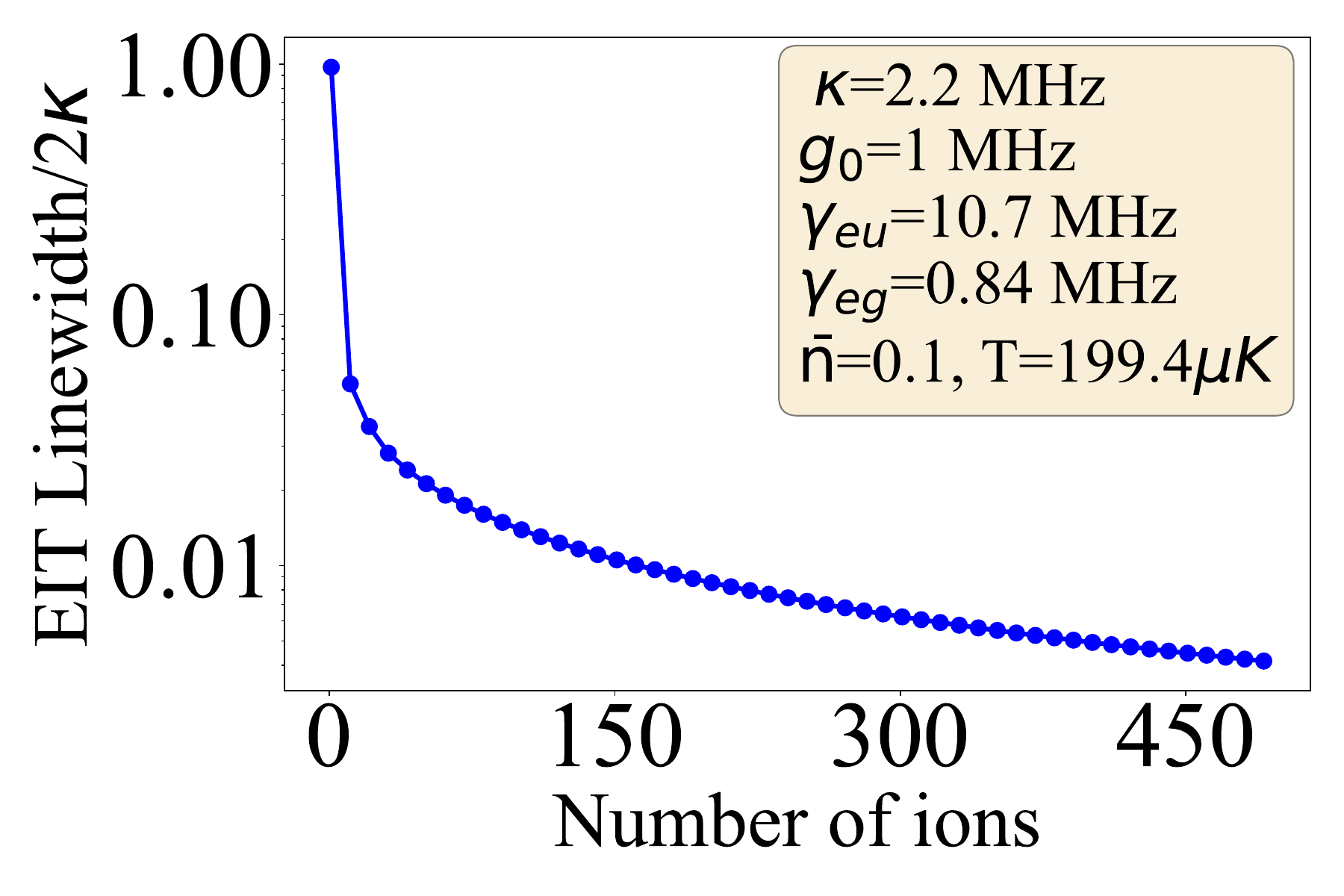}
    \caption{\justifying{Variation of cavity-EIT linewidth with the number of ions with large decay rate of ion excited state. The simulation parameters are shown in the inset.
    }}
    \label{fig:width_vs_nat_high_decay}
\end{figure}

A large decay rate induces decoherence in the system, leading to a negligible cavity-EIT effect, and only the bare cavity signal remains visible at the output. Several groups worldwide have demonstrated the cavity-EIT effect using ion/atom species with a large decay rate from the excited state \cite{PhysRevA.85.023818,mucke2010electromagnetically}. The experimental results from these systems show that for a single ion the cavity-EIT signal is not clearly visible; however, increasing the number of atoms/ions coupled to the cavity mode leads to the re-appearance of cavity-EIT spectrum. This occurs because, although the ion-cavity system is not configured in the strong coupling regime for a single ion/atom (as the decay rates far exceed the single atom cavity coupling factor), the effective coupling strength increases with multiple ions coupled to the intracavity field. Consequently, the collective ion-cavity coupling can overcome the large decay rates of the excited states, as discussed in \autoref{sec:Multi_ion_small_coupling_factor}. More discussion about their system and an excellent agreement of our simulation results with the analytical expression and the reported experimental data can be found in Appendix \ref{appen:a} and \ref{appen:b}.
\newline
\par Motivated by these studies, we extend our ion-thermometry method to cases where the excited state decays rapidly to the ground states in a three-level lambda system. The cavity-EIT peaks become visible when considering a multi-ion system, and the linewidth decreases with the increasing number of atoms, which is the usual behavior similar to that obtained in \autoref{sec:Multi_ion_small_coupling_factor}. The variation of the cavity linewidth with the number of atoms, obtained from simulations performed for up to 500 ions, along with the corresponding simulation parameters, is depicted in \autoref{fig:width_vs_nat_high_decay}.\\
\begin{figure}[htbp]
    \includegraphics[width=\columnwidth]{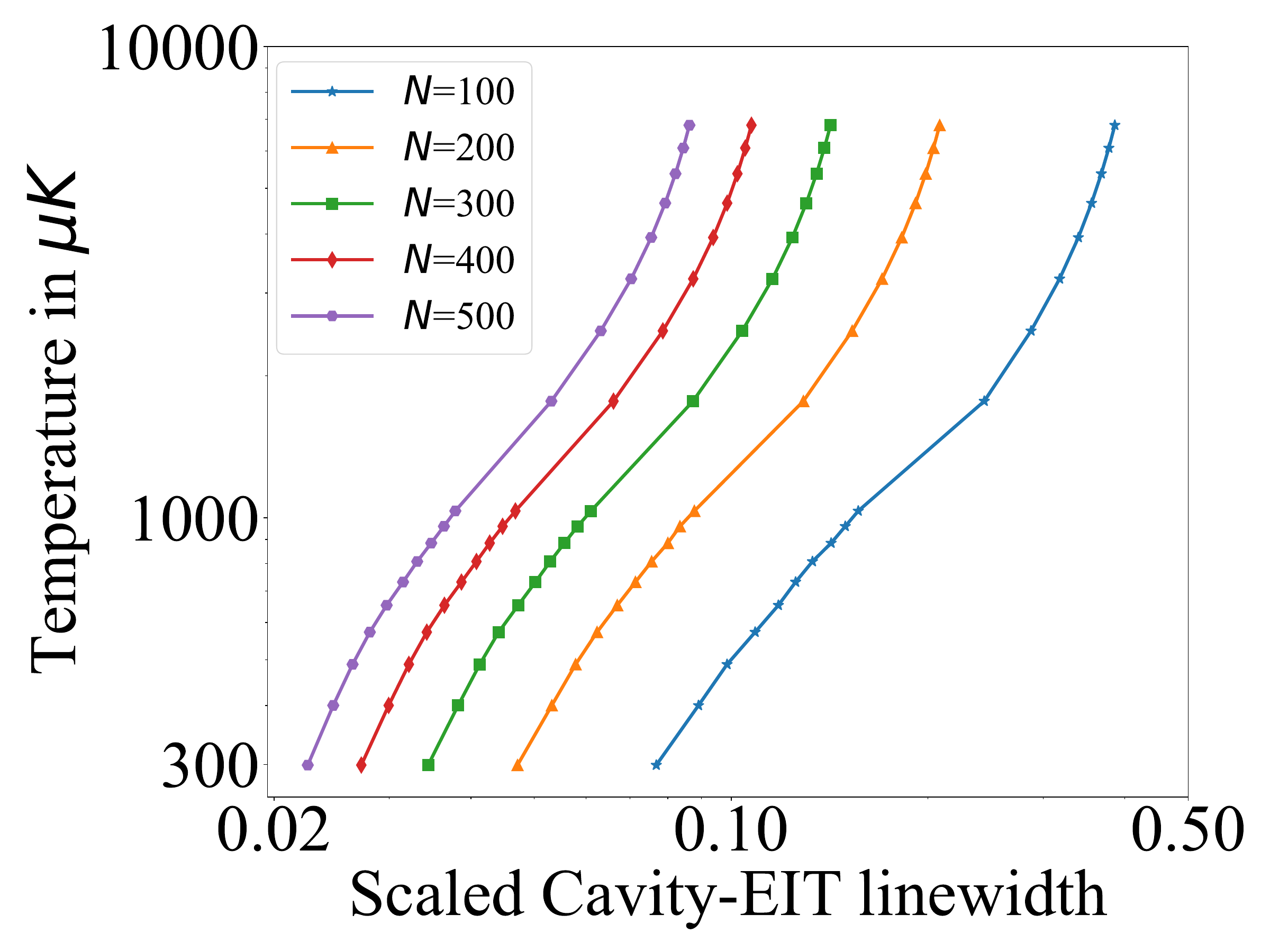}
    \caption{\justifying{
     Temperature vs linewidth for large decay from the excited level. For running the simulation, the cavity decay rate ($\kappa$) is taken as 0.4$MHz$, whereas the cavity coupling factor is taken as $g_{0}=3\kappa$.  The decay rates from excited state $\ket{e}$ to $\ket{u}$ and $\ket{g}$ are $\gamma_{eu}=20\kappa$, $\gamma_{eg}=7\kappa$ respectively. The control field Rabi frequency is chosen as $\Omega_c=2.5\kappa$, whereas the phonon coupling rate to the phonon bath is taken as $\gamma_b=0.5\kappa$. The secular frequency of the ion is taken as $\omega_{sec}=15MHz$. The maximum probe field intensity ($\epsilon / \kappa$) is taken as $0.01$. Here, the cavity-EIT linewidth is scaled with respect to the empty cavity linewidth ($2\kappa$).
    }}
    \label{fig:thermometry_high_decay}
\end{figure}
\par Interestingly, when running the simulations at different temperatures, i.e., for different values of $n_{th}$, the linewidth does not vary at all. This occurs because, for systems with large decay rates, the resolved sideband condition ($\gamma\ll\omega_{sec}$) is no longer satisfied. As a result, the motional sidebands cannot be resolved, which is essential for the thermometry process. Therefore, we conclude that, in addition to strong coupling, a necessary condition for implementing this thermometry method is that the ion must operate in the resolved sideband regime. The only remaining option in such cases is to increase the trap frequency $\omega_c$; in other words, a tighter confinement of the ion with a deeper trapping potential is required.
\newline
 \par To assess the applicability of the thermometry method in the presence of an excited state with a relatively large decay rate ($>$ 7-8 MHz), we analyze the temperature dependence of the cavity-EIT linewidth for different numbers of ions as shown in \autoref{fig:thermometry_high_decay}. A distinct and monotonic dependence of the normalized linewidth on the temperature is preserved even for large decay rates ($\gamma_{eu}=20\kappa$, $\gamma_{eg}=7\kappa$). Although the decoherence caused by spontaneous emission, which also partially broadens the EIT feature, collective enhancement ($g_{eff} \propto \sqrt{N}$) strengthens the effective light-matter interaction to compensate for decoherence. These results demonstrate that the suggested cavity-EIT thermometry method is robust even in regimes of significant radiative decay.

\section{Projected sensitivity of cavity-induced EIT thermometry}\label{sec-section8}
The method of cavity-EIT based ion thermometry works well for an ideal system, where the sidebands are well resolved and ion is strongly coupled to the cavity. Although such a system does-not exist in practice. Accordingly, in the previous two sections, we discussed how this thermometry method is applicable to both the weak-coupling regime and to systems with large excited-state decay rates. It is observed that tight confinement ($\omega_{sec}\gg\gamma$) is absolutely required whereas the strong coupling can be achieved by using low mode volume cavity or using multiple number of atoms.\\
\par An important question that might arise pertains to how accurately our proposed technique can sense variations in temperature and what is the potential sensitivity corresponding to the average phonon occupation. Addressing this question provides insight into the practical applicability of the method across different parameter regimes.  Furthermore, a sensitivity analysis helps to quantify the precision of the thermometry technique and highlights its strengths as well as its limitations. We did not provide any discussion of sensitivity for ideal system as at present its not practical. \\

\par From the observations in \autoref{Linewidth_vs_temp_for different_Nat} and \autoref{fig:thermometry_high_decay}, it is evident that the variation of the linewidth is not uniform across the entire temperature range. Consequently, the sensitivity of our method depends on the specific temperature regime under consideration. To quantify this behavior, we define the temperature sensitivity as the change in temperature corresponding to a unit change in the linewidth, which can be expressed as,
\begin{equation}
    S_T=\frac{\Delta(\Delta\nu)}{\Delta T}
\end{equation} 
where $\Delta\nu$ represents the cavity-EIT linewidth. Similarly the sensitivity for $\bar{n}$ can be written as,

\begin{equation}
    S_{\bar{n}}=\frac{\Delta(\Delta\nu)}{\Delta \bar{n}}
\end{equation} 

\par The minimum achievable temperature of the ion, as well as the separation between the motional energy levels, is inherently determined by the secular frequency of the ion trap. This follows from the fact that the effective temperature corresponding to the ground-state energy of the phonon ladder scales with the secular frequency as $T \sim \frac{\hbar \omega_{sec}}{2k_B}$. Consequently, a higher secular frequency leads to a larger energy spacing between the motional levels and therefore to a higher equivalent temperature scale associated with these levels. The minimum detectabtle temperature change on the other hand depends on the minimum resolvable linewidth change dictated by the mirror reflectivity.\\

\par At higher temperatures, the sensitivity appears to be relatively low, whereas it increases at lower temperatures. To estimate the achievable sensitivity, we assume that the minimum detectable change in the cavity-EIT linewidth is $10~\text{kHz}$.
\newline
\begin{figure}[htbp]
    \includegraphics[width=\columnwidth]{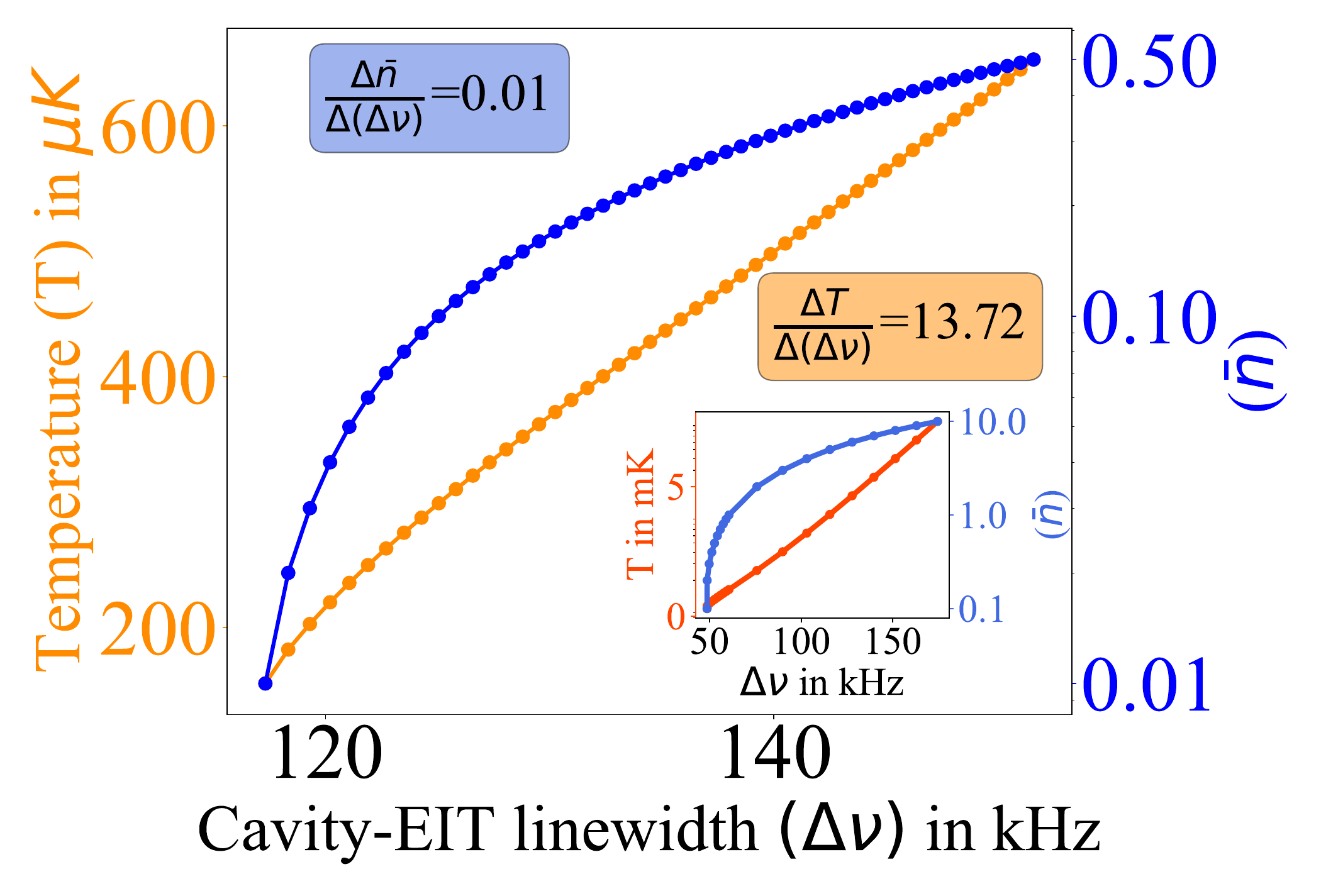}
    \caption{\justifying{Temperature vs cavity-EIT linewidth for systems operating in the strong coupling regime for a single ion.  The parameters are chosen from \cite{PhysRevLett.124.013602}. Single ion cavity coupling $g_{0}=12.3~MHz$, cavity decay rate, $\kappa=4.1~MHz$, excited state decay, $\gamma=12.5~MHz$. The control field Rabi frequency ($\Omega_c$) is taken as $3~MHz$. The secular frequency is taken to be $15~MHz$. The inset plot shows the variation in temperature and average phonon number over a broader range of cavity-EIT linewidth }}
    \label{fig:sensitivity_high_coupling}
\end{figure}
\par We have calculated the sensitivity for both the strong and weak-coupling regimes in the presence of a large excited-state decay rate, which represents a realistic and widely encountered experimental scenario. The parameters used in our simulations are inspired by those reported in the experiments of \cite{PhysRevLett.124.013602} and \cite{albert2011cavity}, with the exception of the secular frequency of the ion. In our case, the secular frequency must exceed the decay rates of the excited state in order to satisfy the resolved sideband condition. Accordingly, the secular frequency is chosen to be approximately $15~\text{MHz}$.
\newline
\par \autoref{fig:sensitivity_high_coupling} represents the single-atom strong-coupling case. Typically, such conditions prevail in fiber cavity-based QED systems. From the slope of the cavity-EIT linewidth versus temperature and $\bar{n}$ plot, we get the value of $14~\mu K/kHz$ and $0.01/kHz$, respectively. But determining $1kHz$ resolution in linewidth can be experimentally challenging, so if we take a minimum of $10~kHz$ change in linewidth, the minimum detectable temperature change will be $140 ~\mu K$ and the minimum detectable deviation in $\bar{n}$ will be $0.1$. As discussed in \autoref{sec:Multi_ion_large_excited_state_decay}, the necessary condition for this thermometry method is to be in the resolved sideband regime ($\gamma\ll\omega_{sec}$). For this reason, the secular frequency is taken as $15~MHz$, which is higher than the excited state decay rate of $12.5~MHz$.
\\
\begin{figure}[htbp]
    \includegraphics[width=\columnwidth]{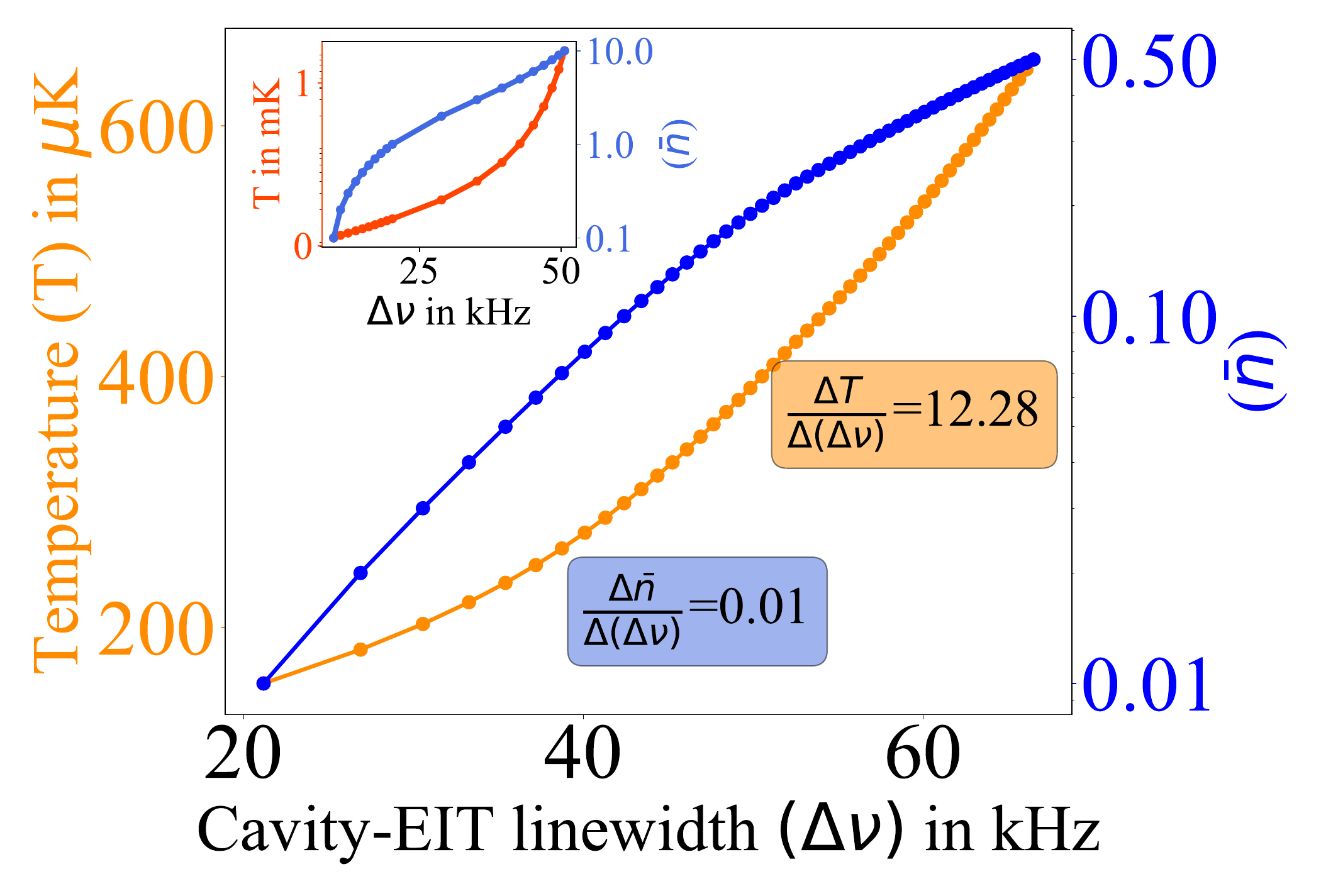}
    \caption{\justifying{Temperature vs cavity-EIT linewidth for systems operating in the weak coupling regime for a single ion. The inset plot is for a higher range of temperature, average phonon number and cavity-EIT linewidth.  The parameters are chosen the same as the system given in \cite{albert2011cavity}. The single ion cavity coupling factor $g_{0}=0.54~MHz$. The cavity decay rate is $\kappa=2.2~MHz$, and the excited state decay rate is $\gamma=12.5~MHz$. The Rabi frequency for control field ($\Omega_c$) is $3~MHz$. The secular frequency is considered as $15~MHz$ to maintain the resolved sideband regime.}}
    \label{fig:sensitivity_low_coupling}
\end{figure}
\par \autoref{fig:sensitivity_low_coupling} illustrates the sensitivity for the multi-atom low-coupling regime. About $500$ atoms are considered to be coupled to the cavity, resulting in effective strong coupling. The sensitivity in this case is about $120~\mu K$ per $10~kHz$ change in linewidth. 
\\
\par The sensitivities in both cases are nearly identical, as both are in an effective strong coupling regime with a similar effective coupling factor. A similar behaviour is also observed when considering the average phonon occupation number ($\bar{n}$).\\

\par In our case, as represented in \autoref{fig:sensitivity_high_coupling} and \autoref{fig:sensitivity_low_coupling}, the assumed secular frequency is ($\sim 15~\mathrm{MHz}$), which is required to satisfy the strict conditions discussed earlier. This secular frequency value is significantly higher than that used in conventional trapped-ion experiments (typically $\sim 1~\mathrm{MHz}$) since the analysis is performed for calcium (\textsuperscript{40}Ca\textsuperscript{+}) ion. As a result, the temperature change corresponding to a given variation in the EIT linewidth, as well as in $\bar{n}$, becomes larger. This reduces the temperature sensitivity of the method, although the sensitivity in terms of the average phonon occupation number $\bar{n}$ remains unaffected.

\section{Roadmap towards experimental realization}\label{sec:roadmap}
We discuss an outline of experimental implementations for the trapped-ion thermometry method discussed in the previous sections of the article. As the thermal state detection method proposed is cavity EIT-based, one needs to engineer a trapped-ion cavity-QED system in the strong-coupling regime ($g_{0}^2\gg \kappa\gamma$). This may be achieved in two ways. The first approach would be to engineer a cavity configuration in the near-concentric or near-confocal regime as per the requirements of the physics objectives. We recommend using a high finesse ($>50000$) cavity, such that the coupling factor ($g_{0}$) is high enough to justify operation in the strong coupling regime ($g_{0}^2\gg \kappa\gamma$) for a single atom/ion coupled to the cavity mode. The second approach focuses on a relatively relaxed regime of multi-ion based strong coupling regime of operation for a low-finesse cavity.\\ 

We need to choose a particular ion species with suitable energy levels, and appropriate ion trap parameters such that the system stays in the resolved sideband regime ($\gamma\ll\omega_{sec}$). Most of the ion species used in cavity-QED experiments globally typically have strong dipole-allowed transitions, with broad linewidths, where sidebands are not resolved in general. Such traps operate with typical trap frequencies in the range ($\omega_{sec}$) $\approx 0.5-2 MHz$. So, one needs a judicious choice of the energy levels, such that the effective three-level system realized has very narrow linewidths for one of the excited levels.\\

One may also consider using a dipole allowed transition (where typical linewidths are in the range of $\approx 5-20 MHz$) for the trapped ion strongly coupled to the cavity field, one should consider operating the ion trap in a very tight confinement (very deep trap potential) of the ion with a secular frequency greater than the transition linewidth ($\Gamma$). Such a mode of ion trap operation has been experimentally validated \cite{PhysRevA.51.3112}, where an \textsuperscript{9}Be\textsuperscript{+} ion was trapped with a very high RF voltage (600 V) and frequency (240 MHz), reaching secular frequencies in the range of $\sim 50~ MHz$ and the sideband transitions were addressed in a dipole-allowed transition.\\

Recent theoretical work has addressed both the challenges and advantages of operating trapped-ion systems at high secular frequencies \cite{rasmusson2026high}. It has shown that high secular frequencies can be achieved through appropriate optimization of the trap design, including the use of high RF drive frequencies (hundreds of MHz), large RF voltages (up to $\sim$1~kV), and suitable electrode geometries. The authors further argue that the associated limitations are primarily technical rather than fundamental and can be overcome with continued advances in modern RF electronics, microfabrication, and trap-design technologies. \\ 

If one considers applying the proposed thermometry method with multiple ions in low- to medium-finesse (finesse values in the range $3000 -15000$) cavities having a relatively  large mode waist ($50-100~\mu m$), about 10 ions coupled to the cavity mode can allow the system to operate in the strong-coupling regime, assuming a single ion cavity coupling factor of $0.5$ MHz and cavity decay rate as well as the excited state decay rate of about $1$ MHz.    

\section{Conclusion}\label{sec:conclusion}
\par We have presented a cavity-assisted thermometry scheme for trapped ions that takes advantage of the dependence of cavity-induced EIT on ions' motional state. Through the explicit integration of thermalization and vibrational sidebands into a cavity-QED $\Lambda$-type system, we show that the mean phonon occupation and the ion temperature can be directly measured by the linewidth of the cavity-EIT transparency window. By means of motion-induced dephasing, thermal phonons perturb the coherence of the EIT dark state, resulting in a systematic expansion of the cavity-EIT resonance. This enables direct extraction of the ion temperature from the measured cavity transmission spectrum, eliminating the need for projective internal-state detection in Rabi spectroscopy. Overall, for the diagnosis of ion temperature in cavity-QED systems, cavity-EIT-based thermometry provides a minimally invasive and experimentally feasible method. The established coupling between ion motion and cavity transmission goes beyond thermometry, offering a platform to investigate measurement-induced effects in cavity-QED systems and enabling continuous monitoring of motional heating

\subsection*{Acknowledgments}
Abhijit Kundu gratefully acknowledges financial support from IIT Tirupati, Government of India. Vijay Bhatt acknowledges support from the NQM (National Quantum Mission), Department of Science and Technology, Ministry of Science and Technology, Government of India through Project No. DST/QTC/NQM/QComm/2024/2(G) administered through the IITM CDOT SAMGYNA TECHNOLOGIES FOUNDATION, IIT Madras, Chennai, Tamil Nadu - 600113. Arijit Sharma acknowledges support from the NQM (National Quantum Mission), Department of Science and Technology, Ministry of Science and Technology, Government of India through Project Nos. DST/QTC/NQM/QComm/2024/2(G) and DST/QTC/NQM/QComm/2024/2(C) administered through the IITM CDOT SAMGYNA TECHNOLOGIES FOUNDATION, IIT Madras, Chennai, Tamil Nadu - 600113.

\subsection*{Conflicts of Interest}

The authors declare no conflicts of interest.

    
    

\bibliography{apssamp}

\appendix
    
\section{Transmission and reflection spectra with a bare cavity}\label{appen:a}
\begin{figure}[htbp]
    \includegraphics[width=\columnwidth]{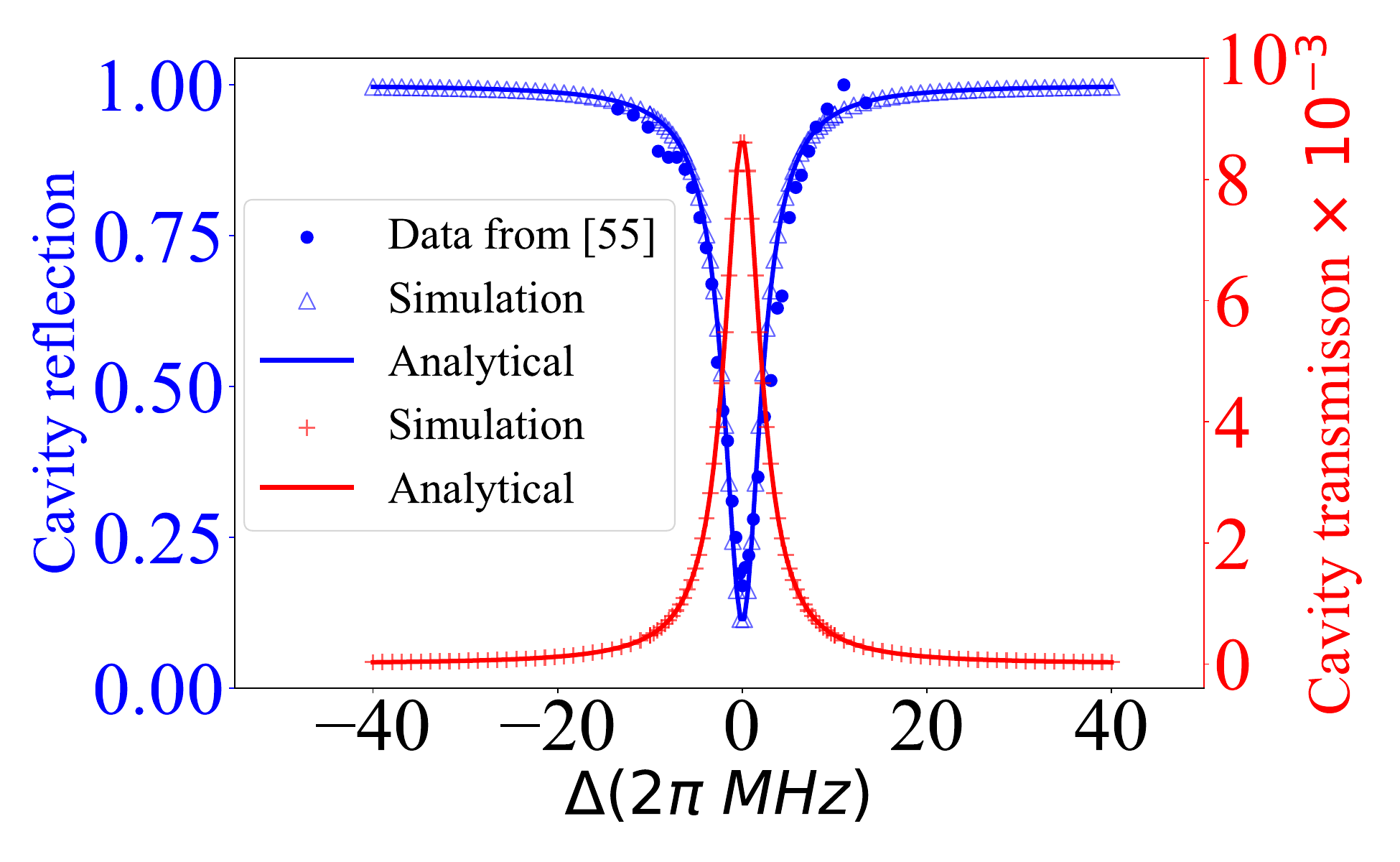}
    \caption{\justifying{
     Cavity reflection from simulation, analytical expression and published experimental data \cite{albert2010light} as a function of probe detuning. The free spectral range of the cavity is $\nu_{FSR} = 12.7~\text{GHz}$. The transmission coefficients of the mirrors are $T_1 = 1500~\text{ppm}$ and $T_2 = 5~\text{ppm}$, while the intracavity loss coefficient is $A = 600~\text{ppm}$. For these parameters, the resulting cavity finesse is approximately $F \sim 3000$.
    }}
    \label{empty_cavity}
\end{figure}

To validate the theoretical model and numerical simulations discussed in the earlier sections of the article, we examine the empty-cavity limit by setting the ion-cavity coupling strength to zero and ignoring all phonon-related contributions in the Hamiltonian. Under these conditions, the system reduces to a driven, dissipative optical cavity described solely by the bare cavity Hamiltonian and the decay channels associated with it.\\

The Hamiltonian of the bare cavity system can be expressed as,
\begin{equation}
H_{\text{bare}}= \Delta_{c} a^{\dagger}a + i\sqrt{2\kappa_{1}}(a^{\dagger}a_{in} - a_{in}^{\dagger}a),
\end{equation}

where $a^{\dagger}$ ($a$) is the creation (annihilation) operator for the cavity mode, $\Delta_{c}$ represents the probe-cavity detuning, and $a_{in}$ is the input field injected into the cavity.\\

Standard input-output relations are used to calculate the resulting transmission ($\mathcal{T}$) and reflection ($\mathcal{R}$) spectra. For an empty cavity, the response follows the well-known Lorentzian resonance profile determined by the cavity decay rates.

\begin{equation}
\mathcal{R} = \left|\frac{a_{\mathrm{ref}}}{a_{\mathrm{in}}}\right|^2 
= \frac{(\kappa - 2\kappa_1)^2 + \Delta_c^2}{\kappa^2 + \Delta_c^2},
\end{equation}

\begin{equation}
\mathcal{T} = \left|\frac{a_{\mathrm{trans}}}{a_{\mathrm{in}}}\right|^2 
= \frac{4\kappa_1 \kappa_2}{\kappa^2 + \Delta_c^2},
\end{equation}

where, $\kappa$ corresponds to the total decay rate of the cavity field, while $\kappa_{1}$ and $\kappa_{2}$ represent the contribution to the cavity loss arising from the imperfect reflectivity of the two cavity mirrors.\\

\autoref{empty_cavity} shows the cavity transmission and reflection spectra obtained from our numerical simulation together with the corresponding analytical expression and previously reported experimental data \cite{albert2010light}. The obtained spectra exhibit the expected Lorentzian lineshape with normalization, reproducing the well-known analytical response of a bare cavity. The validity of the reduced Hamiltonian reflecting the response of the bare cavity and the numerical method used throughout the manuscript are confirmed by the excellent agreement with the standard bare-cavity reflection and transmission profiles. The bare cavity response therefore serves as a baseline reference against which the modifications arising from ion-cavity interactions and phonon-induced effects can be reliably identified in the main results.

\section{Comparison of simulation results with analytical expressions and experimental data}\label{appen:b}

We verify our numerical simulations for a multi-ion cavity-QED system exhibiting a low single-ion cavity coupling factor as well as using ions with a large decay rate of the excited state. These systems can be driven in the collective strong coupling regime by appropriate choice of ions coupled to the cavity mode. We compare the simulated normalized cavity emission spectrum with previously published experimental data and results and compare them with available analytical expressions mentioned in \cite{dissertation_mucke, albert2010light}.\\

We first consider the cavity-EIT model described in \cite{dissertation_mucke}, where a three-level $\Lambda$-type atom interacts with a single mode of an optical cavity. The atomic system consists of an excited state $\ket{e}$ and two long-lived ground states $\ket{g}$ and $\ket{u}$. The cavity mode is coupled to the transition $\ket{g} \leftrightarrow \ket{e}$ with coupling strength $g_0$ while a classical control field with Rabi frequency $\Omega_{c}$ drives the transition $\ket{u} \leftrightarrow \ket{e}$. The cavity mode is driven by a weak probe field with amplitude $\epsilon$ and the cavity output spectrum is measured as a function of the probe-cavity detuning. In the rotating frame, the system Hamiltonian can be expressed as,

\begin{equation}
\begin{aligned}
H = \hbar \Big[
&\Delta_1 \sigma_{ee}
+ (\Delta_1 - \Delta_2)\sigma_{uu}
+ g_{0}(a^\dagger\sigma_{ge} + \sigma_{eg}a) \\
&+ \frac{\Omega_c}{2}(\sigma_{ue} + \sigma_{eu})
+ \epsilon(a^\dagger + a)
- \Delta a^\dagger a
\Big],
\end{aligned}
\end{equation}

where detunings are defined as $\Delta_1 = \omega_{eg} - \omega_p$, $\Delta_2 = \omega_{eu} - \omega_c$ and $\Delta = \omega_c - \omega_p$. $\omega_p$ and $\omega_c$ correspond to the probe and control field 
frequencies, respectively.\\

\begin{figure}[htbp]
    \includegraphics[width=\columnwidth]{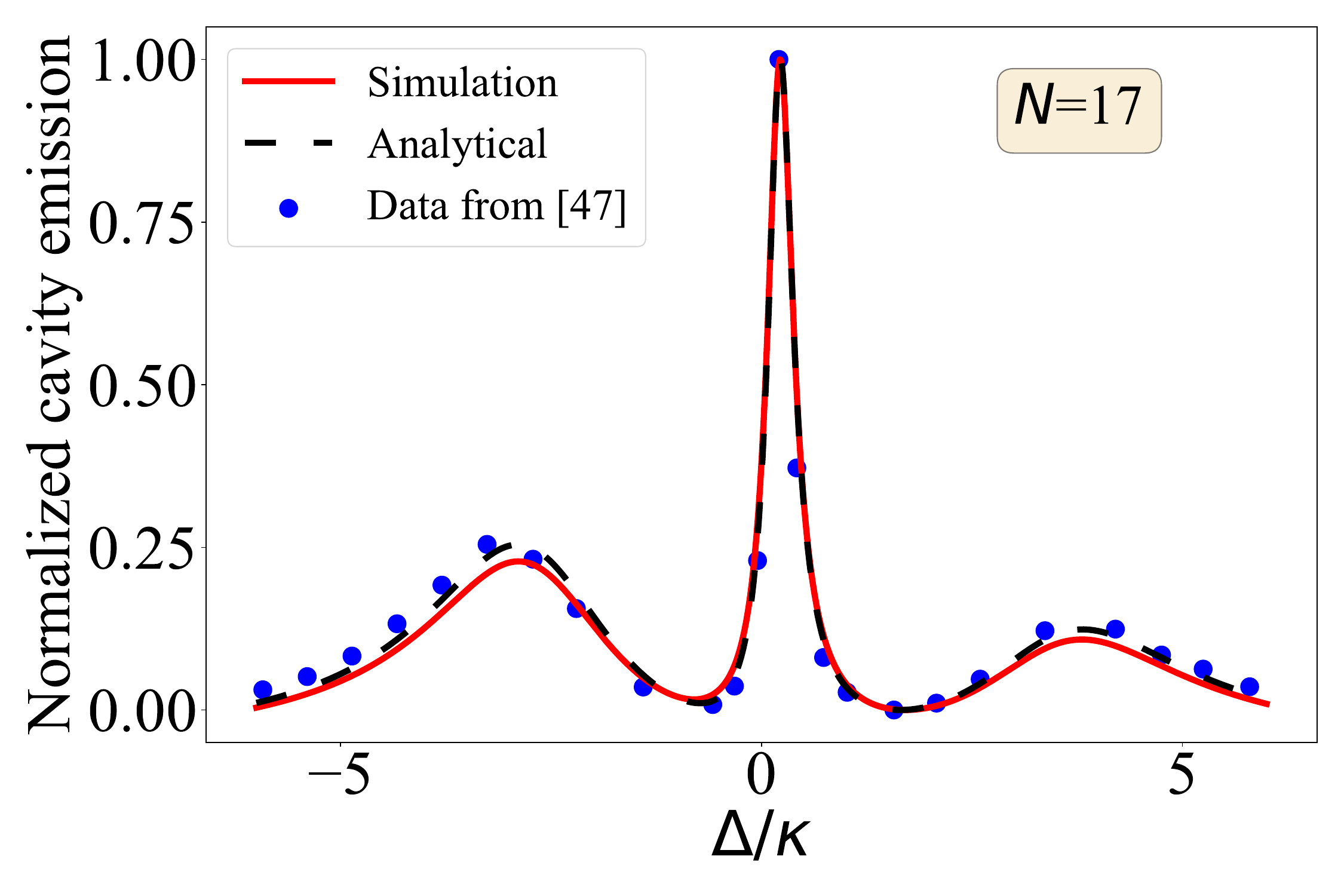}
    \caption{\justifying{
     Normalized cavity emission from simulation, analytical expression and published experimental data \cite{dissertation_mucke} as a function of probe detuning. The cavity decay rate ($\kappa$), is 1 MHz, while the single-atom cavity coupling rate ($g_0$) is $0.7\kappa$. The control field Rabi frequency $(\Omega_c) $ is $3MHz$. The detuning for probe laser ($\Delta_1$) and control laser ($\Delta_2$) are $0.8\kappa$ and $0.55\kappa$ respectively. 
    }}
    \label{simu_exp_analytical_comp}
\end{figure}

The intracavity photon number can be obtained analytically from the steady-state cavity field in the weak-probe limit as,

\begin{equation}
\langle n \rangle = \langle a^\dagger a \rangle
= \frac{|\epsilon|^2}{\left|(\Delta + i\kappa) - \chi \right|^2},
\end{equation}

where $\kappa$ represents the cavity decay rate and $\chi$ is the effective susceptibility of the EIT medium,

\begin{equation}
\chi =
g^{2}N
\frac{\Delta-\Delta_{1}+\Delta_{2}+i\gamma_{u,\mathrm{deph}}}
{\begin{aligned}
&\left(\Delta-\Delta_{1}
+i(\gamma_{eg}+\gamma_{eu})\right)\\
&\times
\left(\Delta-\Delta_{1}
+\Delta_{2}
+i\gamma_{u,\mathrm{deph}}\right)
-\frac{\Omega_c^{2}}{4}
\end{aligned}}
\label{eq:susceptibility}
\end{equation}
With these analytical expression and the parameters listed in \cite{dissertation_mucke} we compute the transmission spectrum and compared it with the outcomes of our numerical model. \autoref{simu_exp_analytical_comp} highlighted the normalized cavity emission as a function of probe detuning $(\Delta /\kappa)$ having $N=17$. As observed, the simulation accurately reproduces the key spectral features, including the central transparency peak and two sidebands.\\

The accuracy of the theoretical implementation is confirmed by the excellent agreement between the simulations and the analytical prediction. Furthermore, the close overlap with experimental data confirms the physical assumptions incorporated in the model.This agreement shows that the numerical framework reproduces experimentally observed behavior with high reliability and captures the cavity-EIT dynamics.\\

\begin{figure}[htbp]
    \includegraphics[width=\columnwidth]{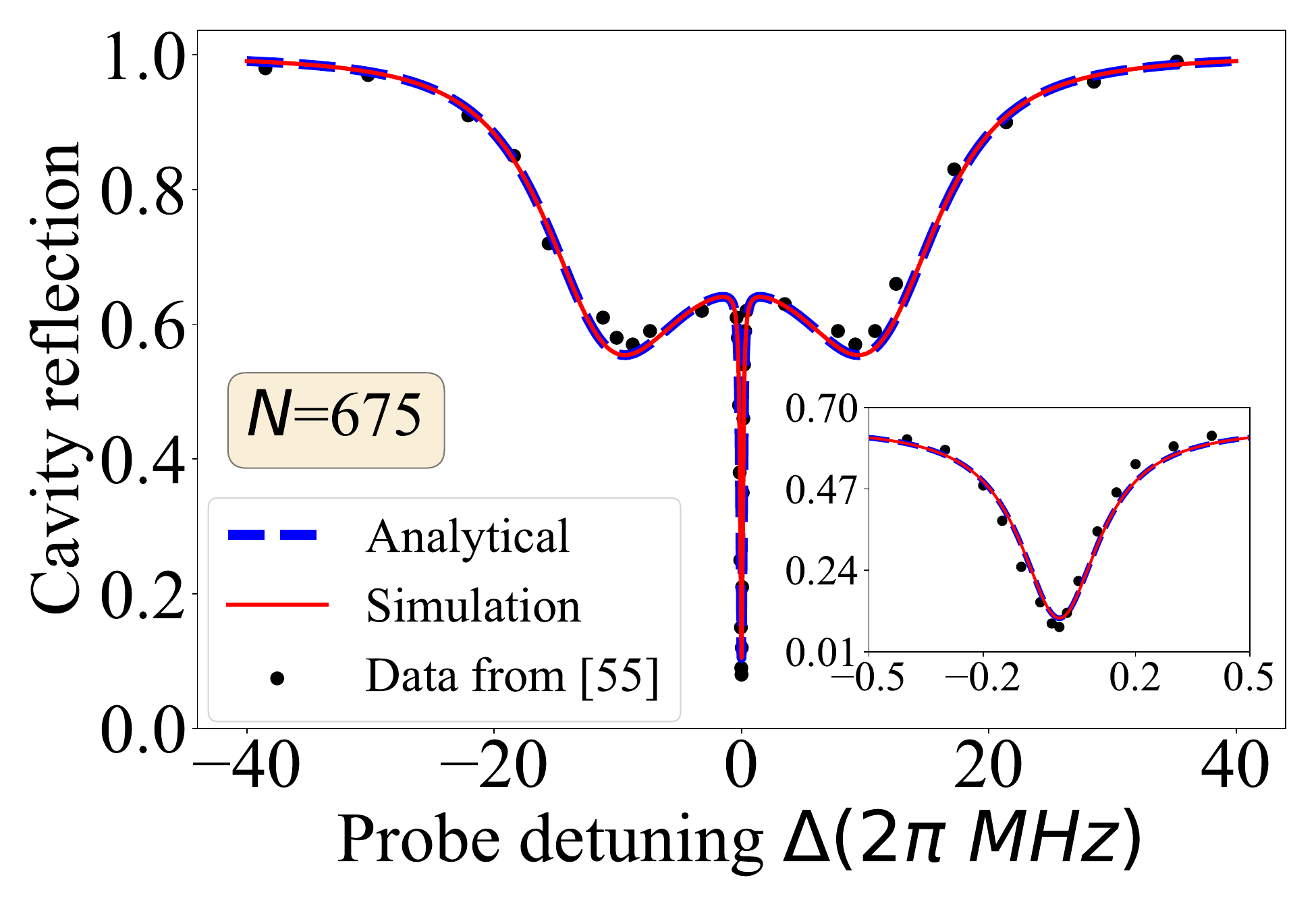}
    \caption{\justifying{
     Cavity reflection from simulation, analytical expression and published experimental data \cite{albert2010light} as a function of probe detuning. The decay rate for 1st mirror ($\kappa_1$), 2nd mirror ($\kappa_2$) are $1.6 MHz$ and $0.5 kHz$ respectively. The decay rates from excited state to first and 2nd ground state are, $\gamma_{31}=12.6 ~MHz$ and $\gamma_{21}=0.6 ~kHz$ respectively. The single atom coupling factor $g_{0}=0.48~MHz$, while control field Rabi frequency, $\Omega_c=4.1 ~ MHz$.
    }}
    \label{albert_reflection}
\end{figure}

To further validate our numerical model, we also compare our results with the reflection spectra published in \cite{albert2010light}. This work examines the cavity reflection from an ensemble of atoms coupled to the cavity mode in contrast to the previous work in \cite{dissertation_mucke}, which concentrates on the cavity transmission spectrum. Specifically, for the system of about 500 atoms, the reflection spectrum was investigated analytically and experimentally in  \cite{albert2010light}.\\ 

The cavity-EIT spectrum shown in \autoref{albert_reflection} is deduced from the analytical expression of the normalized cavity reflectivity with respect to the input field, as explained in \cite{albert2010light}. The reflectivity can be written as,

\begin{equation}
\begin{aligned}
R &= \left|{\frac{a_{ref}}{a_{in}}}\right|^2 \\
&=
\frac{
\left[\kappa_1 - \kappa_2 - \kappa_A - \mathrm{Im}(\chi_{\Lambda})\right]^2
+
\left[\Delta - \mathrm{Re}(\chi_{\Lambda})\right]^2
}{
\kappa^2 + \Delta^2 + \left|\chi_{\Lambda}\right|^2
+ 2\left[\kappa\,\mathrm{Im}(\chi_{\Lambda}) - \Delta\,\mathrm{Re}(\chi_{\Lambda})\right]
}
\end{aligned}
\label{eq:albert_reflection}
\end{equation}
Here $a_{ref}$ and $a_{in}$ are the reflected and input field amplitude respectively. $\kappa_1$, $\kappa_2$ and $\kappa_A$ are the decay rates for field losses from the cavity first mirror, second mirror and additional internal losses due to various mechanisms. $\Delta$ is the probe field detuning and $\kappa$ is the total decay rate ($\kappa=\kappa_1+\kappa_2+\kappa_A$). $\chi_{\Lambda}$ is the three-level non-linear susceptibility of the atomic ensemble, which can be expressed as,
\begin{equation}
\chi_{\Lambda} =
\frac{i g_{0}^{2} N}{\gamma + i\Delta}
\frac{\ln(1+s)}{s}
\label{eq:chi_lambda}
\end{equation}

Where, $g_{0}$ is the single ion coupling factor, $N$ is the effective number of atoms coupled with the cavity field, $\gamma$ is the combined decay rates for the excited state. The saturation parameter for the two-photon transition is $s$, which can be articulated as,

\begin{equation}
s =
\frac{\Omega_C^2 / 2}{(\gamma_{12} + i\delta)(\gamma + i\Delta)}
\label{eq:s_parameter}
\end{equation}

Where, $\Omega_{c}$, is the control field Rabi frequency, $\gamma_{12}$ is the dephasing rate between the ground states. $\delta$ is the two-photon detuning, which the difference between probe laser detuning ($\Delta$) and control laser detuning ($\Delta_{32}$), $\delta=\Delta-\Delta_{32}$. 

\section{Dependency of cavity-EIT linewidth on cavity coupling parameter 
\texorpdfstring{$g_0$}{g} and control field Rabi frequency 
\texorpdfstring{$\Omega_{c}$}{Omega_c} for different 
\texorpdfstring{$n_{th}$}{n_th}} \label{appen:c}

\begin{figure}[htbp]
    \centering

    \begin{subfigure}[t]{0.49\linewidth}
        \centering
        \includegraphics[width=\linewidth]{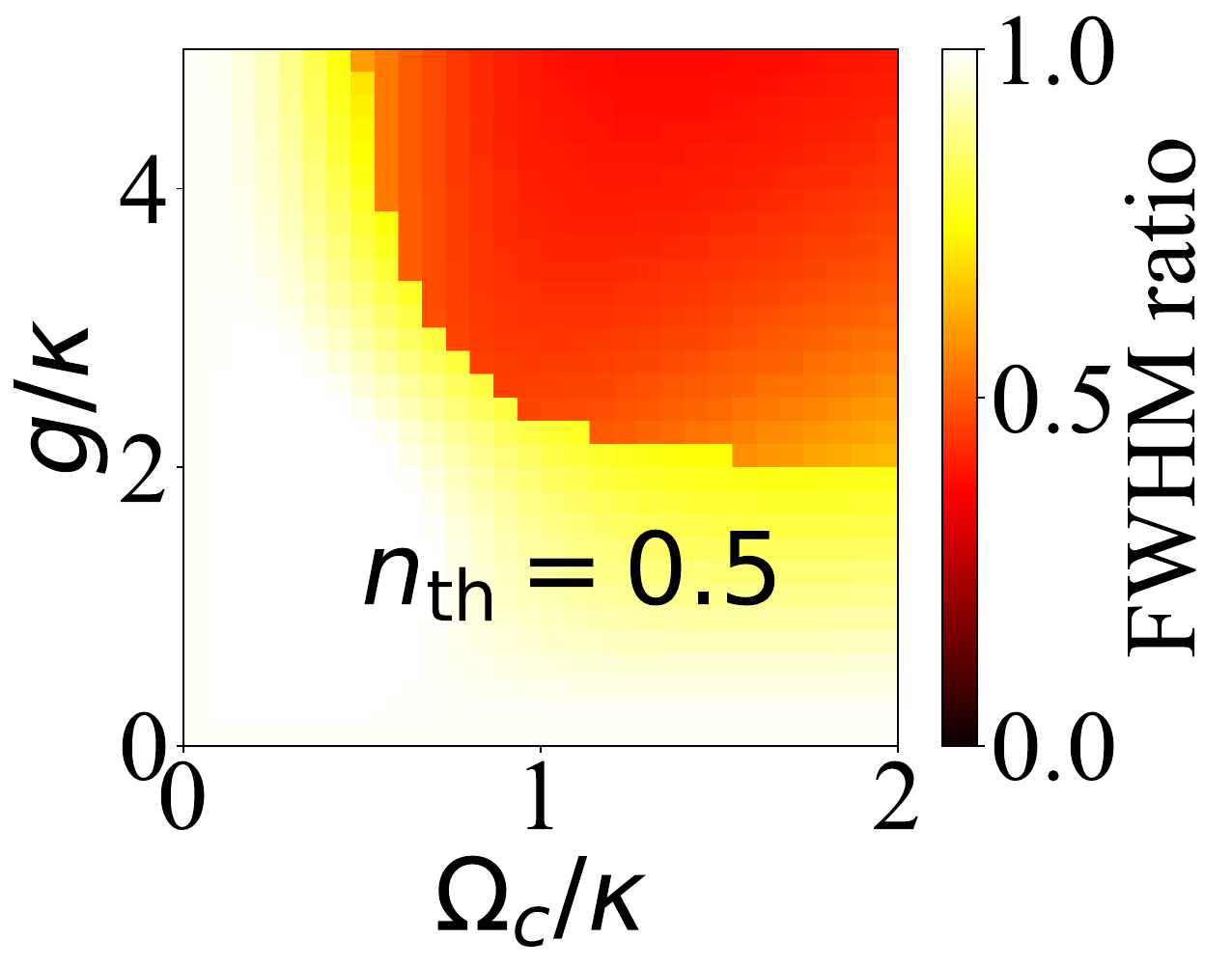}
        \caption{}
        \label{fig:4a_maps_a}
    \end{subfigure}
    \hfill
    \begin{subfigure}[t]{0.49\linewidth}
        \centering
        \includegraphics[width=\linewidth]{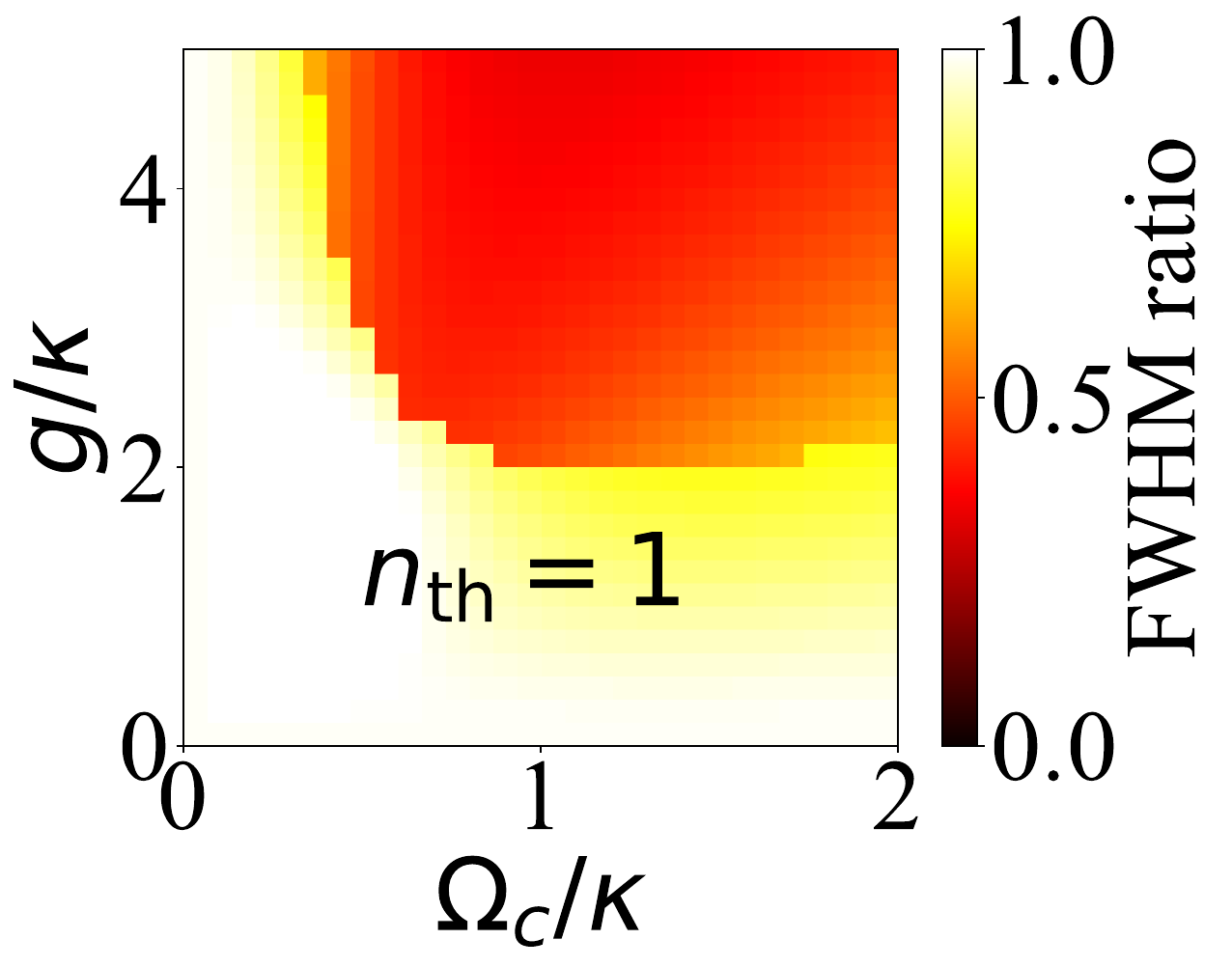}
        \caption{}
        \label{fig:4b_maps_b}
    \end{subfigure}

    \vspace{0.2cm}

    \begin{subfigure}[t]{0.49\linewidth}
        \centering
        \includegraphics[width=\linewidth]{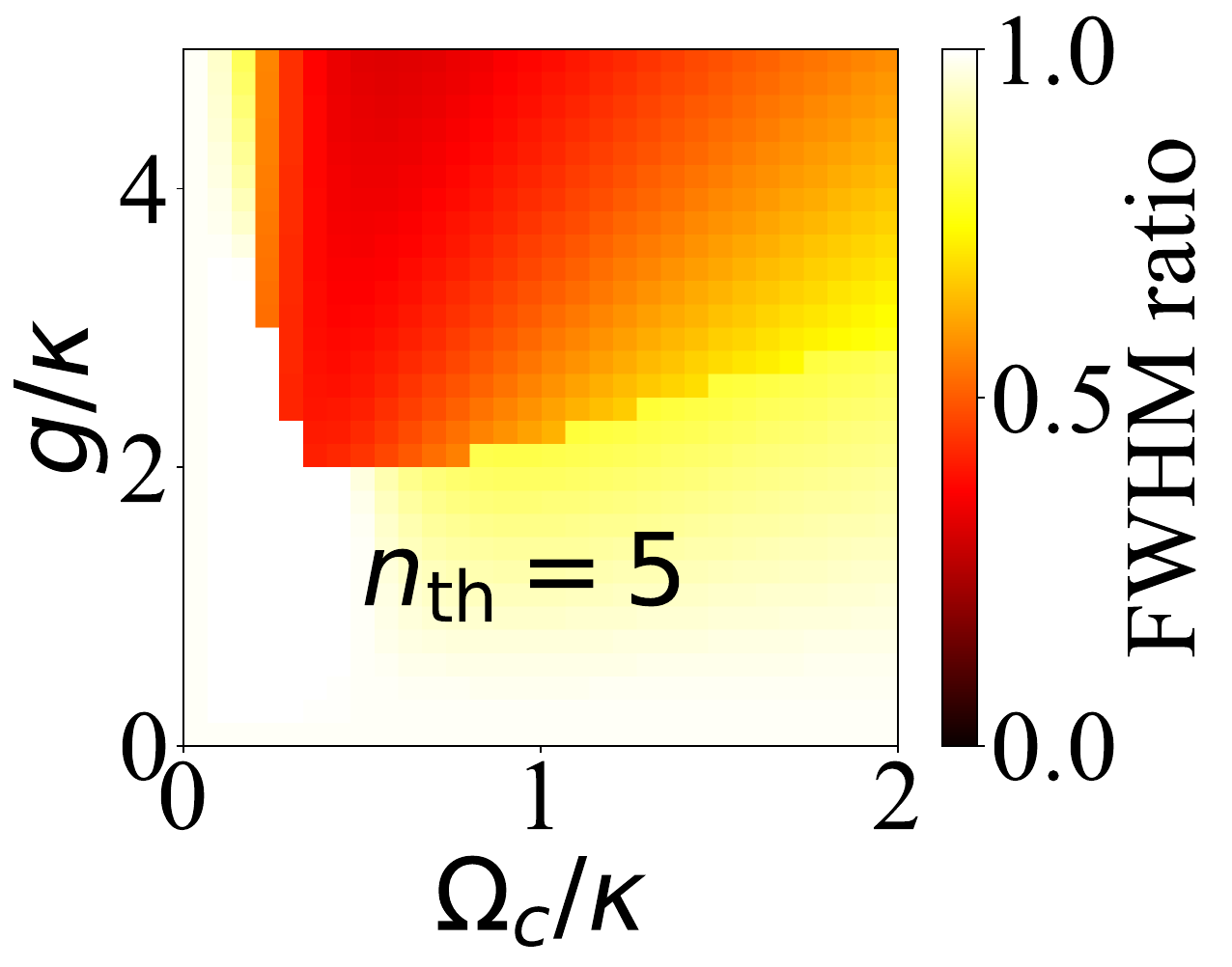}
        \caption{}
        \label{fig:4c_maps_c}
    \end{subfigure}
    \hfill
    \begin{subfigure}[t]{0.49\linewidth}
        \centering
        \includegraphics[width=\linewidth]{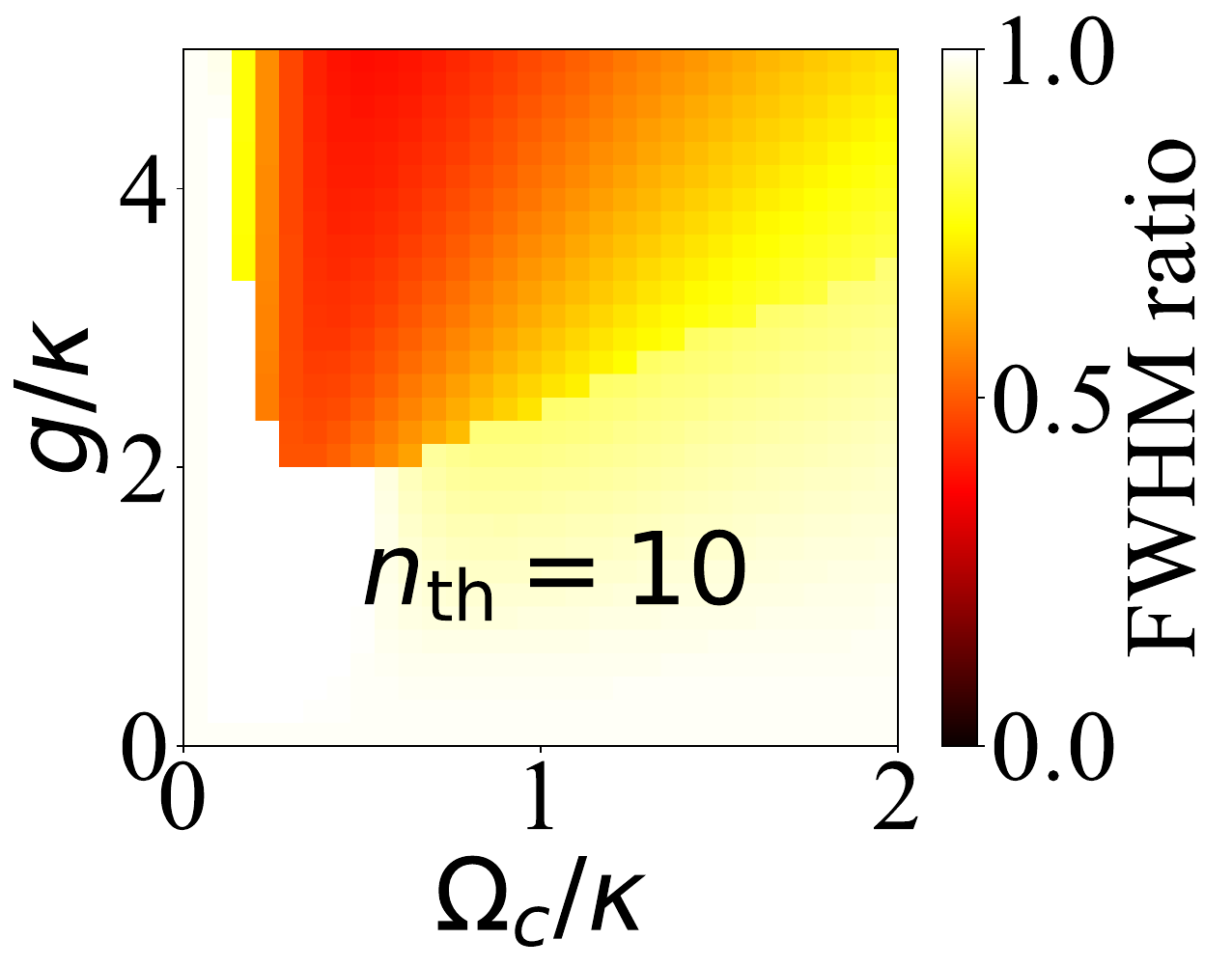}
        \caption{}
        \label{fig:4d_maps_d}
    \end{subfigure}

    \caption{\justifying{
    Two-dimensional maps of the cavity-EIT linewidth as a function of the
    atom--cavity coupling strength $g_0$ and the coupling-laser Rabi frequency
    $\Omega_c$ for different mean thermal phonon numbers
    $\bar{n}_{\mathrm{th}}$.
    }}
    \label{fig:2d_maps}
\end{figure}

 A two-dimensional map of the cavity-EIT linewidth represented as the normalized full width at half maxima (FWHM) ratio as a function of the normalized control field Rabi frequency $\Omega_{c}$ and the normalized atom-cavity coupling strength $g_{0}$ for various mean thermal phonon numbers $n_{th}$ is shown in \autoref{fig:2d_maps} For a weakly excited motional state ($n_{th}=0.5$) (see \autoref{fig:4a_maps_a}), the linewidth ratio remains close to unity over a large region of the parameter space. This suggests that thermal motion only slightly alters EIT coherence. As the value of thermal occupation increases $n_{th}=1$ (see \autoref{fig:4b_maps_b}), a clear reduction of the FWHM ratio occurs, particularly in the regime of strong ion-cavity coupling and weak control fields. This reduction signals the onset of thermally induced decoherence. A higher phonon number ($n_{th}= 5, 10$) (see \autoref{fig:4c_maps_c}, \autoref{fig:4d_maps_d}) shows a more noticeable trend in which the parameter space is dominated by extended regions of reduced linewidth ratio. In this regime, thermal motion significantly perturbs the ion-cavity interaction, leading to enhanced dephasing and a degradation of the EIT interference. 
As the control-field strength increases, it partially mitigates this effect by stabilizing the dark state against motional fluctuations. The strong sensitivity of the cavity response to the ions' motional state is highlighted by the systematic and temperature-dependent modification of the cavity-EIT linewidth shown in these plots. This provides a solid foundation for thermometry using cavity-based EIT.

\section{Verification of different sideband transition -related phenomenon }\label{appen:d}

As proof of principle of our approach, we performed verification of different sideband transition related characteristics. In the case of sideband transitions, a two-level system dressed with vibrational levels is sufficient to address different properties. The sideband Hamiltonians used here are consistent with the Lamb–Dicke-expanded interaction employed in the main text. The two kinds of sideband transitions (BSB and RSB) are described by the Hamiltonian which is similar to the 4th term (with the multiplication of the Lamb-Dicke parameter $\eta$) of \autoref{eq:Ham_with_phonon}  with the reduced Hilbert space of only photon and phonon. The Hamiltonian for blue and red sideband transition can be written as \cite{leibfried2003quantum},
    \begin{equation}
    \begin{aligned}
        H_{BSB}=\eta\Omega_{c}\left(\sigma_{eu} b^{\dagger} + \sigma_{ue} b\right)\\
        H_{RSB}=\eta\Omega_{c}\left(\sigma_{eu} b + \sigma_{ue} b^{\dagger}\right)
        \end{aligned}
        \label{eq:Ham_sideband}
    \end{equation}

For a thermal distribution of ion  the strengths of red and blue sideband transitions are unequal and depends on the phonon number ($n$). The probability of occupying the excited state for RSB and BSB transition is related by the formula \cite{PhysRevA.61.063418},
\begin{equation}
    \frac{P_{RSB}}{P_{BSB}}=\frac{n}{n+1}
    \label{eq:P_red_P_blue_ratio}
\end{equation}
\begin{figure}[htbp]
    \Centering

    \begin{subfigure}[t]{0.49\linewidth}
        
        \includegraphics[width=\linewidth]{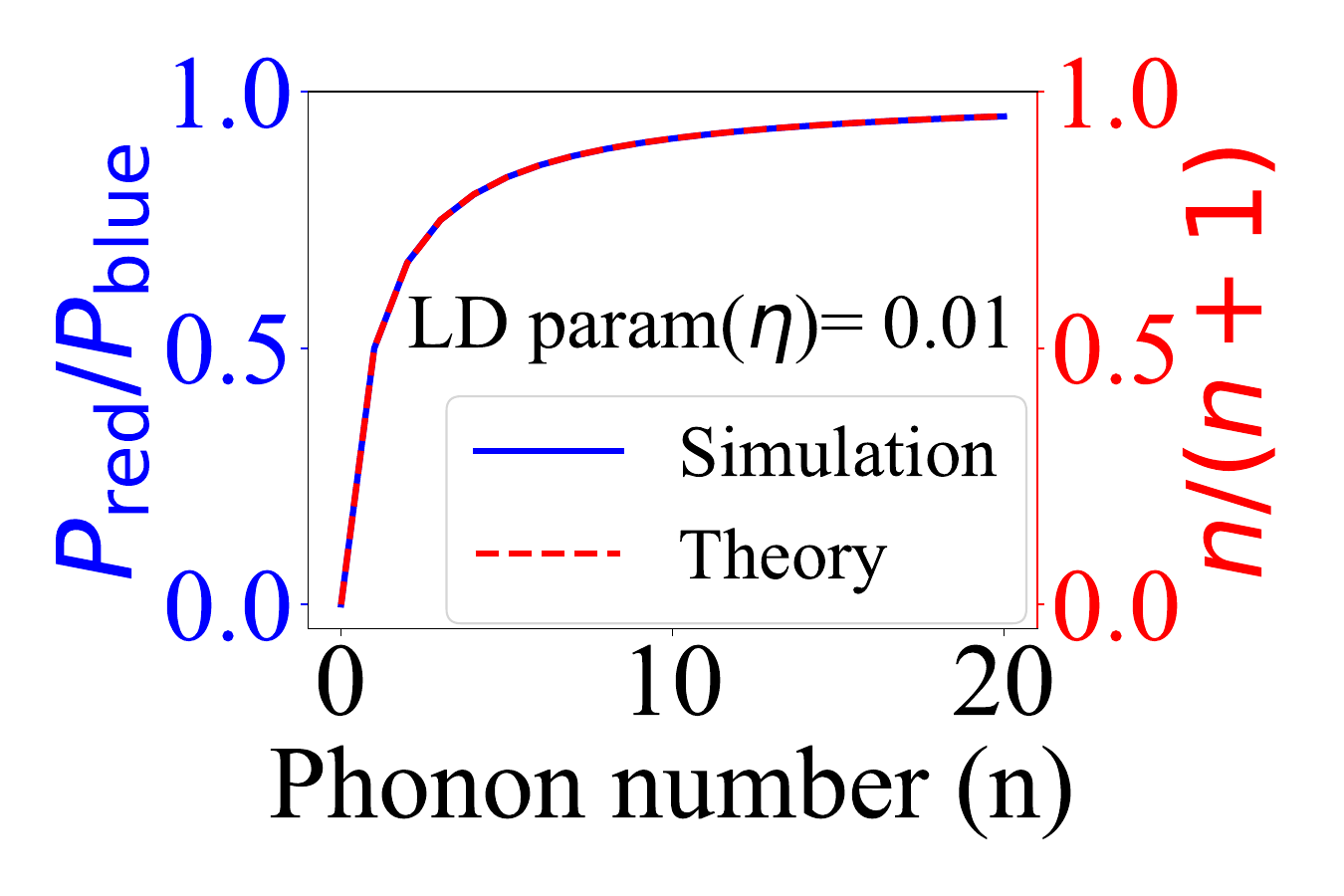}
        \caption{}
        \label{fig:PredPblue_a}
    \end{subfigure}
    \begin{subfigure}[t]{0.49\linewidth}
        
        \includegraphics[width=\linewidth]{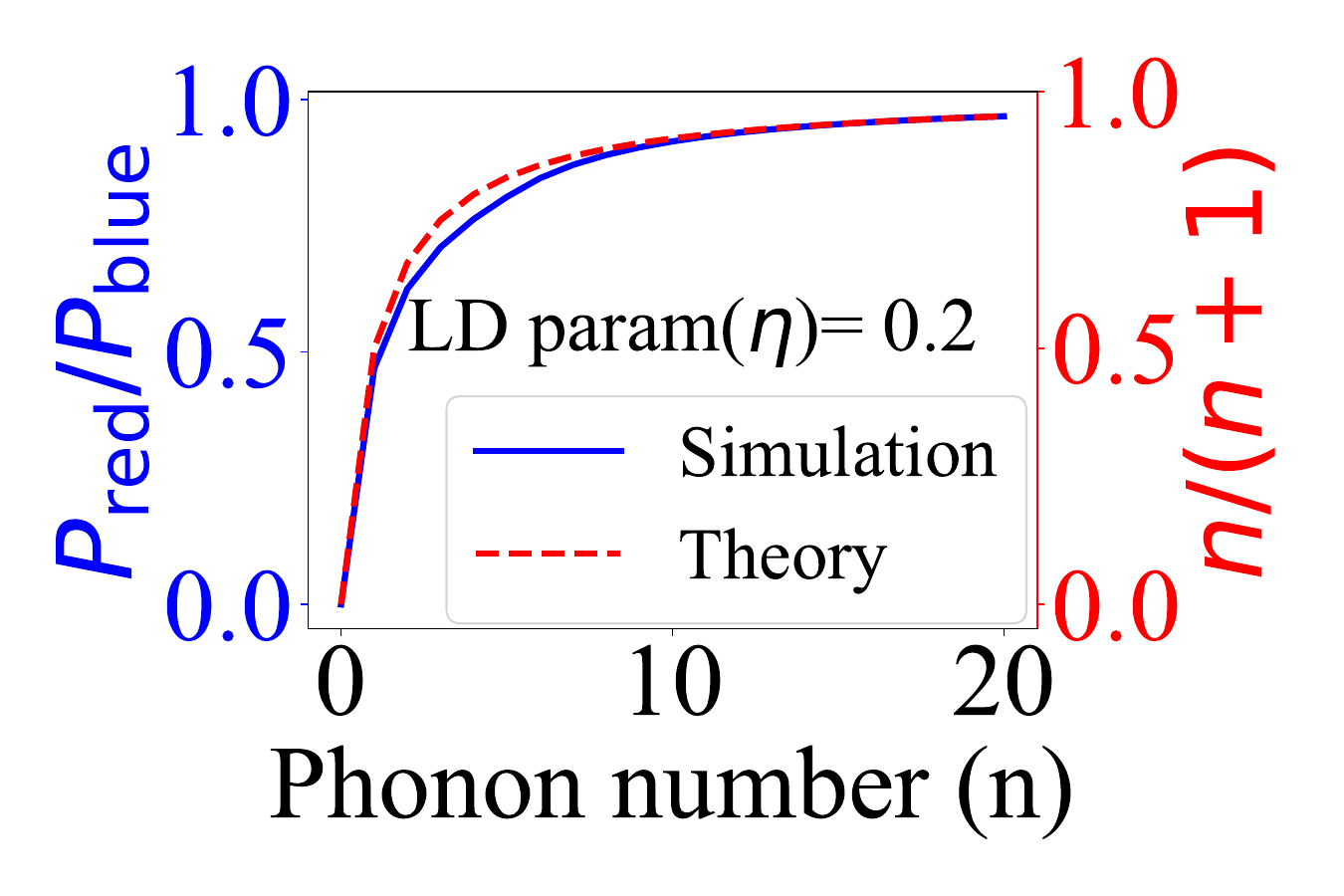}
        \caption{}
        \label{fig:PredPblue_b}
    \end{subfigure}


    \begin{subfigure}[t]{0.49\linewidth}
        
        \includegraphics[width=\linewidth]{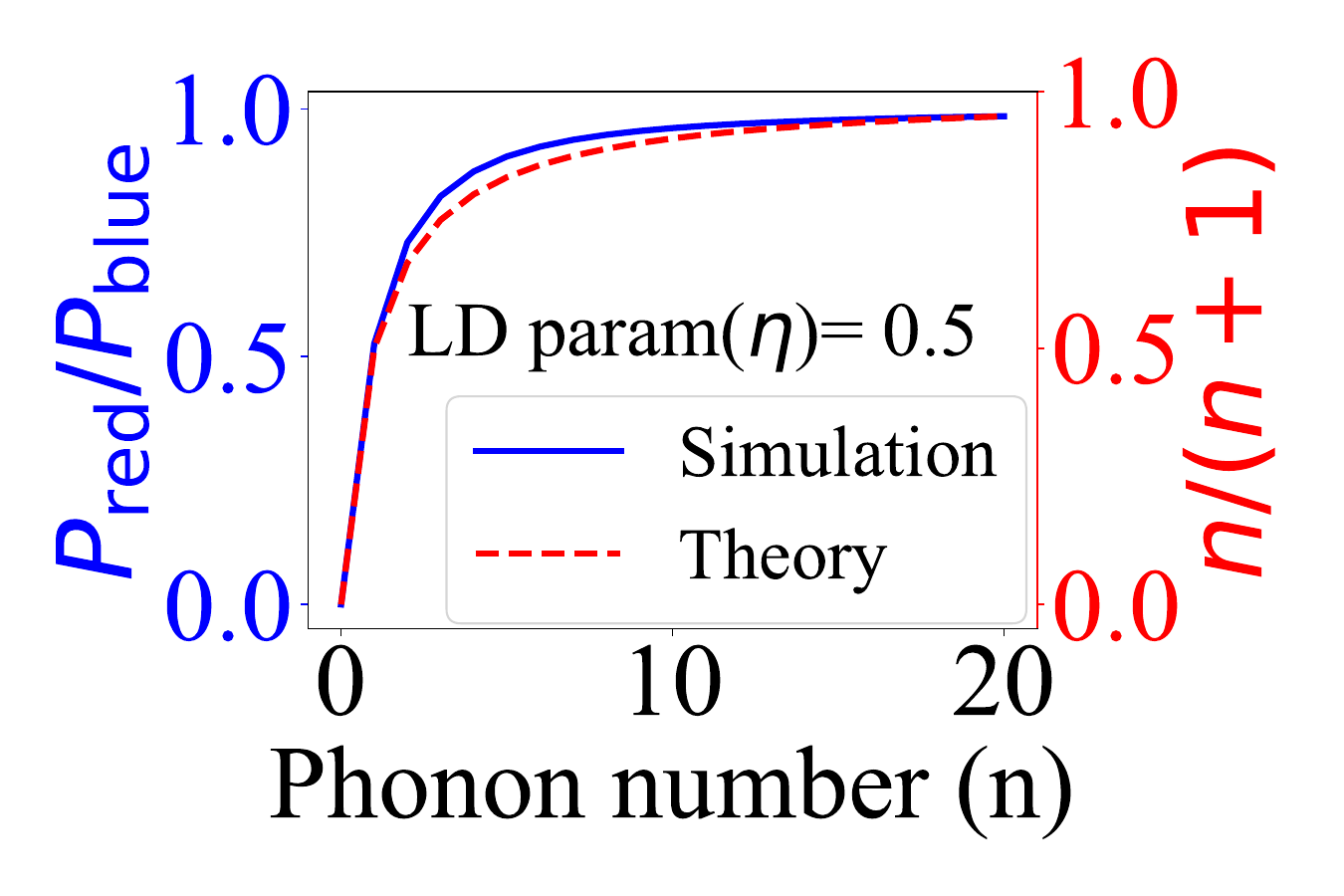}
        \caption{}
        \label{fig:PredPblue_c}
    \end{subfigure}
    \begin{subfigure}[t]{0.49\linewidth}
        
        \includegraphics[width=\linewidth]{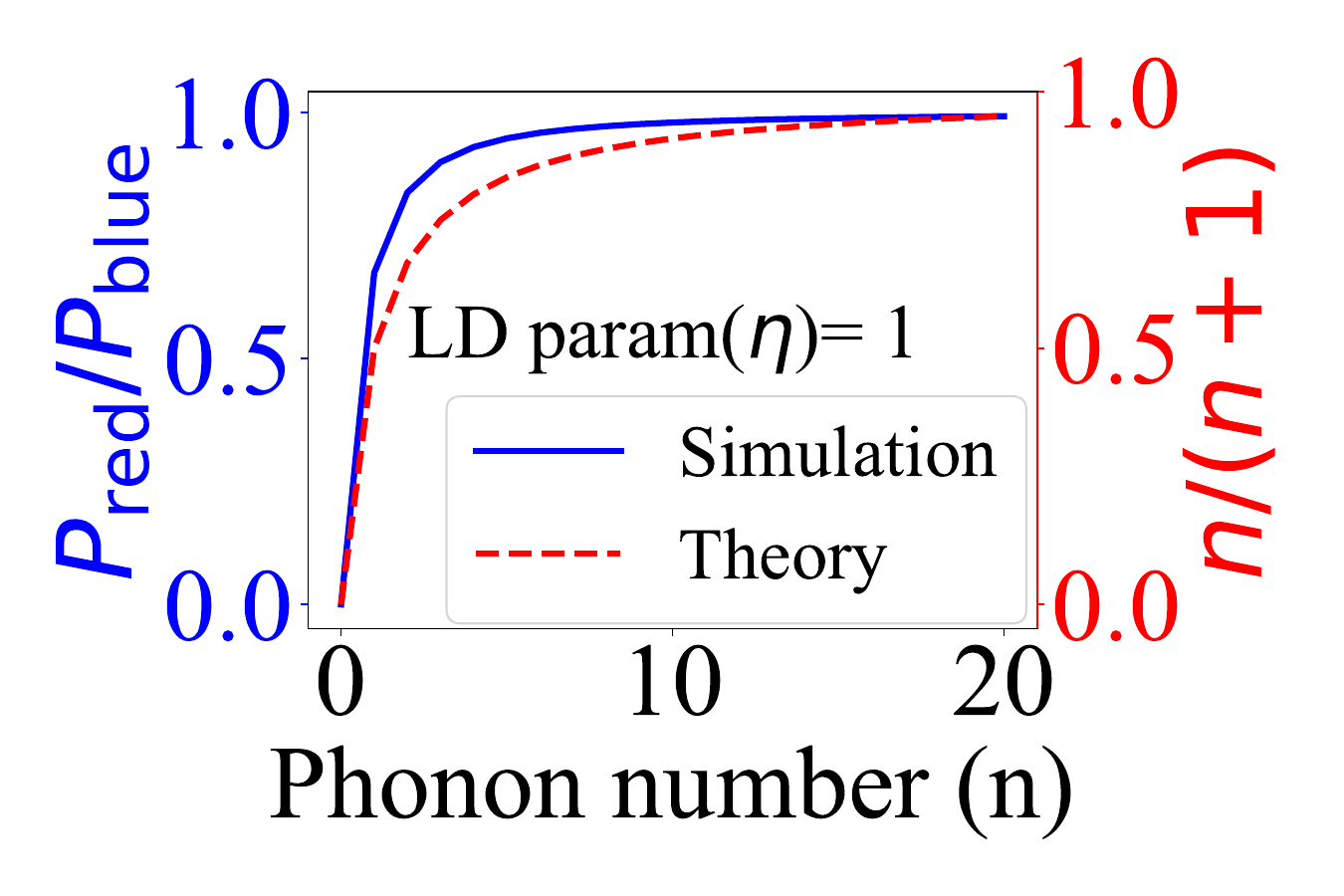}
        \caption{}
        \label{fig:PredPblue_d}
    \end{subfigure}

    \caption{\justifying{Ratio of excited state occupation probability for blue and red sideband transition is compared with phonon number ratio. Plots (a)--(d) correspond to increasing values of the Lamb--Dicke parameter $\eta$.}}

    \label{fig:P_red_P_blue}
\end{figure}
We verified this principle for a range of phonon numbers and different Lamb-Dicke parameters. In \autoref{fig:P_red_P_blue}, the comparison between the probability ratio from simulation and phonon ratio from \autoref{eq:P_red_P_blue_ratio} has been shown for four different Lamb-Dicke parameters. At lower values of Lamb-Dicke parameters, these two quantities are well-matched, which consistent with the theory. For higher Lamb-Dicke parameters the simulation result and theory start to deviate as the boundary of the Lamb-Dicke regime ($\eta \ll 1$) is approached, whereas sideband transitions assume that the ion is well within the Lamb-Dicke regime. 

\begin{figure}[htbp]
\begin{subfigure}[t]{0.49\linewidth}
\includegraphics[width=\linewidth]{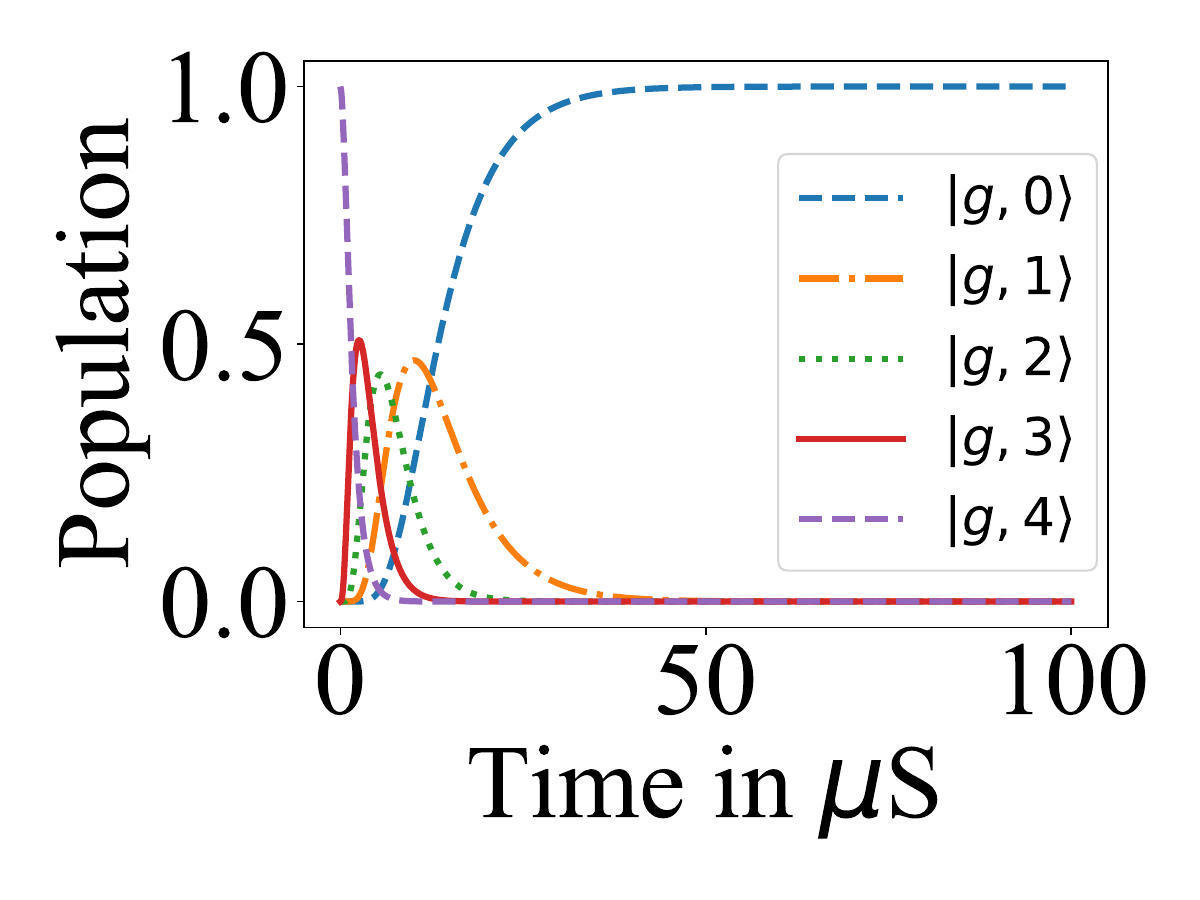}
\caption{}
\label{fig:state_evo_sideband_a}
\end{subfigure}
\hfill
\begin{subfigure}[t]{0.49\linewidth}
\includegraphics[width=\linewidth]{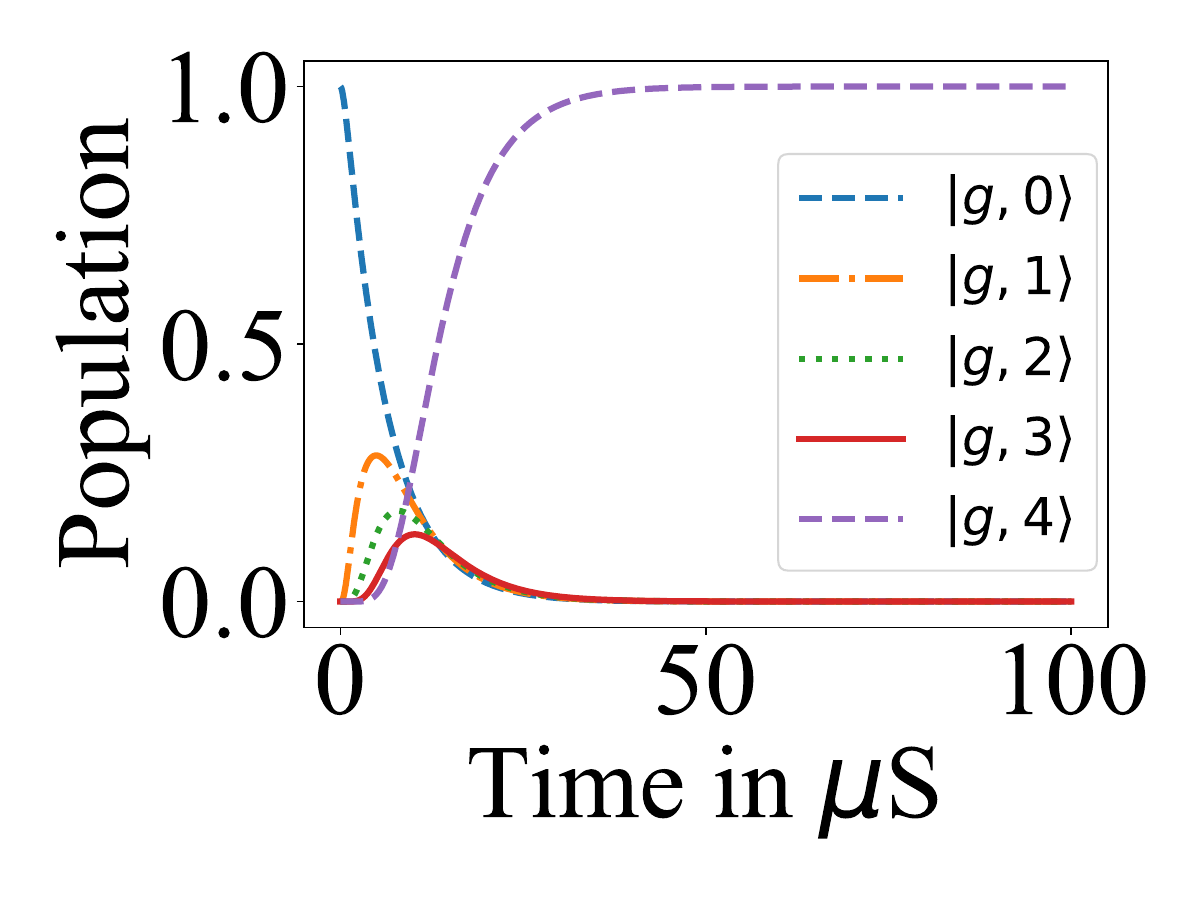}
\caption{}
\label{fig:state_evo_sideband_b}
\end{subfigure}
\caption{\justifying {Simulated state evolution with time. The simulation runs with Lamb-Dicke parameter, $\eta=0.2$, Rabi frequency, $ \Omega=2MHz$, and excited state decay rate, $\gamma=2\Omega$.(a) Time evolution of different states in red sideband transition. The initial state is taken as $|g,5\rangle$, after applying sequential RSB pulses, phonon number decreases, and finally the system reaches the ground state, $|g,0\rangle$.(b) Time evolution for blue sideband transition. population goes from initial $|g,0\rangle$ to higher phonon state $|g,5\rangle$.}
}
\label{fig:state_evo_sideband}
\end{figure}
A larger application of the sideband transition is the cooling of the atom to its motional ground state \cite{PhysRevLett.62.403,PhysRevA.60.439}. When the resolved sideband regime is obtained ($\gamma\ll\omega_{sec}$) and the laser is tuned to a RSB transition, the atom loses one quanta of energy upon excitation ($\ket{g,n}\rightarrow \ket{e,n-1}$). From the excited state, there is a high chance that it spontaneously decays to the ground state with the same motional quanta ($\ket{g,n-1}$). Thus, a motional quanta has been reduced in this process. By repeating this process multiple times, the vibrational quanta has been reduced in each steps and eventually we reached to the ground state of the atom (see \autoref{fig:state_evo_sideband_a}). Similarly, if the laser is tuned to the BSB transition, starting from a ground state, the atom lands in a higher phonon state depending on the number of sideband pulses applied (see \autoref{fig:state_evo_sideband_b}).




\end{document}